\documentclass[12pt,a4paper]{article}

\usepackage[small,bf,hang]{caption}
\usepackage[left=2.5cm,right=2.5cm,top=2.6cm,bottom=2.5cm]{geometry}
\usepackage[T1]{fontenc}
\usepackage[colorlinks,linktocpage]{hyperref}

\usepackage{amsmath,amssymb,amsthm,amsfonts,bigints,float,tensor,braket,dsfont,bm,mathrsfs,cite,setspace,enumerate,graphicx,xcolor}

\definecolor{rood}{RGB}{230,36,0}
\definecolor{groen}{RGB}{55,210,0}
\hypersetup{citecolor=groen, linkcolor=rood}

\DeclareMathAlphabet{\mathpzc}{OT1}{pzc}{m}{it}
\DeclareMathOperator{\tr}{tr}

\newcommand{\n}{\\[3mm]}
\newcommand{\ts}{\textstyle}

\newcommand*\tageq{\refstepcounter{equation}\tag{\theequation}}

\newcommand{\pri}{\null^{\prime}}
\newcommand{\prisq}{\null^{\prime\raisebox{1pt}{$\scriptstyle{2}$}}}

\usepackage[normalem]{ulem}

\newcommand{\ip}[2]{\left\langle{#1}\middle|{#2}\right\rangle}                  
\newcommand{\out}[1]{\left|{#1}\middle\rangle\middle\langle{#1}\right|}                  
\newcommand{\bk}[3]{\left\langle{#1}\middle|{#2}\middle|{#3}\right\rangle}      
\newcommand{\kb}[2]{\left|{#1}\middle\rangle\middle\langle{#2}\right|}      
\newcommand{\exx}[1]{\left\langle{#1}\right\rangle}                             

\newcommand{\op}[1]{\hat{#1}}

\renewcommand{\epsilon}{\varepsilon}
\newcommand{\w}{\omega}

\usepackage{dsfont}
\newcommand{\id}{\mathds{1}} 
\newcommand{\ninfty}{{-\infty}}

\newcommand{\hint}{H^\text{int}}

\newcommand{\ssub}[1]{{\protect\scalebox{0.9}{$_#1$}}}

\newcommand{\I}{\ssub{I}}
\newcommand{\G}{\ssub{G}}

\newcommand{\D}{\mathcal{D}}
\renewcommand{\O}{\mathcal{O}}

\newcommand{\vrp}{\vec{r}\, \pri}

\newcommand{\vr}{\vec{r}}

\newcommand{\uarr}{\scalebox{0.9}{$\uparrow$}}
\newcommand{\darr}{\scalebox{0.9}{$\downarrow$}}
\newcommand{\K}{\mathbf{K}}

\begin{document}
\renewcommand{\baselinestretch}{1.10}\normalsize

\title{How does interference fall?}

\author{Patrick J. Orlando\footnote{patrick.james.orlando@gmail.com}, 
Felix A. Pollock\footnote{felix.pollock@monash.edu}, 
Kavan Modi\footnote{kavan.modi@monash.edu}\\
{\small \it School of Physics and Astronomy, Monash University, Victoria 3800, Australia}}

\maketitle
 
\pagenumbering{arabic}

{\small

\begin{center}
{\normalfont\scshape \large Abstract}
\end{center}
\begin{quotation}
We study how single- and double-slit interference patterns fall in the presence of gravity. First, we demonstrate that universality of free fall still holds in this case, \emph{i.e.}, interference patterns fall just like classical objects. Next, we explore lowest order relativistic effects in the Newtonian regime by employing a recent quantum formalism which treats mass as an operator. This leads to interactions between non-degenerate internal degrees of freedom (like spin in an external magnetic field) and external degrees of freedom (like position). Based on these effects, we present an unusual phenomenon, in which a falling double slit interference pattern periodically decoheres and recoheres. The oscillations in the visibility of this interference occur due to correlations built up between spin and position. Finally, we connect the interference visibility revivals with non-Markovian quantum dynamics.
\end{quotation}}

\newpage

Since the days of Galileo and Newton, it has been known that acceleration under the influence of gravity is independent of an object's mass~\cite{Galilei:1953ua, Newton:821668}. This peculiarity has led to the proposition of various gravitational equivalence principles which, if broken, represent a departure from our current understanding of the theory of gravity. Einstein's theory of general relativity is fundamentally classical, describing gravity on large length scales in terms of curvature of the underlying spacetime metric. Although it is possible to formulate quantum field theories on a static curved metric, it remains unclear how existing theory should be modified to describe gravity on the quantum mechanical scale~\cite{birrell1984quantum}. Whilst the work we present here does not attempt to quantise gravity, it demonstrates that there is much insight to be gained from exploring non-relativistic quantum mechanics in weak-field gravity.

In the weak-field limit, a Newtonian description of gravity provides a satisfactory approximation and is, advantageously, compatible with the Hamiltonian formulation of quantum mechanics; however, its disadvantage lies in the concealment of relativistic effects, such as gravitational time dilation and the gravitational redshift of photons. Fortunately, one need not utilise the complete machinery of general relativity to take these effects into account. In fact, lowest order relativistic effects can be introduced by simply considering the mass contributions of different energy states, as given by the mass-energy relation $E=mc^2$ of special relativity~\cite{Einstein:1911wk}. 

This is true even in the case of internal energy and becomes particularly interesting for quantum systems, whose internal energy can exist in superposition. Recent work by Zych and Brukner~\cite{Zych:2015vm} treats this by promoting mass to an operator, the purpose of which is to account for the effective mass of quantised internal energy. In addition to introducing lowest order relativistic effects, this construction provides a new quantum mechanical generalisation of the Einstein equivalence principle to superpositions of energy eigenstates.

The role that Newtonian gravity plays in quantum theory was perhaps best highlighted by the famous experiment of Colella, Overhauser and Werner (COW), who demonstrated interference of cold neutrons due to a relative phase acquired due to the difference in gravitational potential between two arms of an interferometer. We include details of the COW experiment in Appendix~\ref{sec:COW}.

More recently, the theory of ultra-cold atom condensates has provided a way to test gravitational equivalence principles with quantum systems, by using optically trapped atomic gases as an integrated interferometer \cite{Inguscio:1602444, Berrada:2013bn,Peters:1999iz,Bonnin2015, Zhou2015}. The short de~Broglie wavelength of an atom makes atomic interferometers highly sensitive, whilst the macroscopic nature of the condensate allows for a high degree of control. Proposals for tests on board the international space station have been put forward which, if performed, are expected to surpass the best classical tests by a factor of 100~\cite{QTEST2015}. Finally, tests of the uniquely quantum mechanical equivalence principle for superpositions have also been proposed~\cite{PJORLANDO2015}.

In this article, we study how single- and double-slit interference patterns fall due to gravity. Initially, we ignore the lowest order relativistic effects introduced by internal degrees of freedom and find (unsurprisingly) that the interference patterns fall just as classical objects do; in other words, the universality of free fall holds for spatially delocalised quantum systems. 

We then pedagogically introduce the mass operator and use it to explore non-Newtonian effects on quantum systems with quantised internal energy. One such system is a particle with intrinsic spin incident on a double slit in a gravitation field. We demonstrate that when placed in a uniform magnetic field, the internal energy results in periodic decoherence and re-coherence of the double-slit pattern. This result is an example of decoherence due to gravitational time~dilation presented by Pikovski \emph{et al.} \cite{Pikovski:2015du} and other related works \cite{Zych:2011jz, Zych:2012kq}. 

The decoherence occurs due to the buildup of correlations between the spin and position degrees of the particle. We identify the oscillations in the visibility of the interference fringes as a signature of non-Markovian quantum dynamics \cite{arXiv:1512.00589}, and demonstrate explicitly how memory effects play a role in the evolution of these fringes. This illustrates that the tools of open quantum systems theory can help us clearly understand Newtonian gravity in a quantum mechanical context.

\section*{Dropping a quantum interference pattern} 

\begin{figure}[t]
\begin{center}
\includegraphics[width=0.6\textwidth]{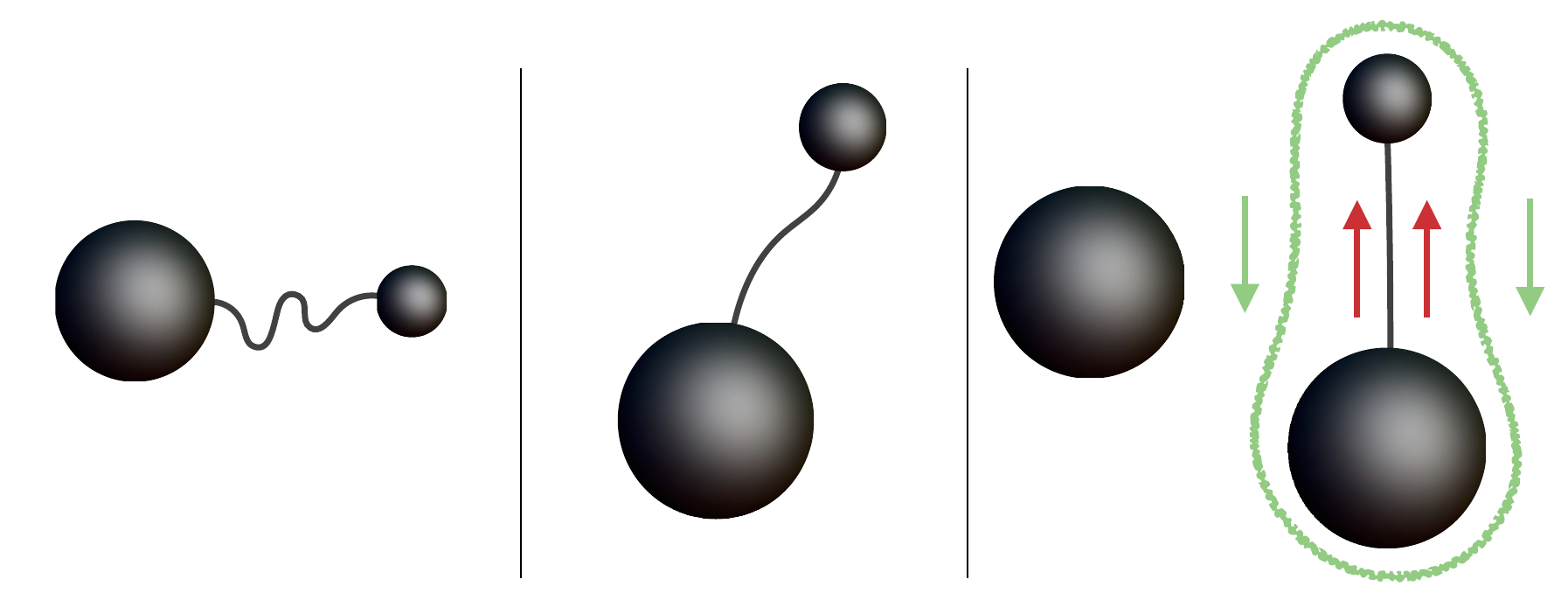}
\caption{{\bf Galileo's thought experiment.} Galileo considered three spheres composed of the same material. Two of the spheres had identical mass, whilst the third sphere was much lighter. He then imagined attaching a rope between the small mass and one of the larger masses, and wondered what would happen if all three were simultaneously dropped from the leaning tower of Pisa. According to Aristotle, the small mass should fall slower than the large mass, pulling the rope taught and impeding the acceleration of the larger mass. One would then expect to see the solitary large mass hit the ground before the attached pair. However, one could also consider the pair of attached masses as a single body, whose mass exceeds that of the large mass alone. In this case, the attached pair of masses would be expected to hit the ground before the solitary large mass. This results in a logical contradiction, from which the only escape is to conclude that both the small and large masses fall with the same acceleration.}
\label{fig:galileo}
\end{center}
\end{figure}

General relativity arose from the concept that gravitational effects are a result of the underlying spacetime geometry. Whilst three fundamental forces of nature: electromagnetism, the strong force and the weak force; all depend on the internal properties of matter, gravity, in the Newtonian regime, depends only on the mass of the particle. Further, its dependence on the mass is such that the dynamics are completely independent of the particle itself. This is often attributed to Galileo in a famous thought experiment, devised to refute Aristotle's claim that the gravitational acceleration of a body is proportional to its mass. His very elegant thought experiment, described in Figure~\ref{fig:galileo}, led to the conclusion that all objects must fall at the same rate, regardless of their mass. This is known as the \emph{universality of free fall}, and has profound consequences for theories of gravity.

In this section we study how quantum interference patterns fall due to gravity. We imagine that massive quantum particles (say neutrons) are ejected towards a single or double slit. Once the particle passes through the slit, it falls freely under the influence of gravity, while simultaneously interfering with itself.

Here, we are concerned with the possibility of interesting gravitational effects appearing in a single slit diffraction or double slit interference experiment. In accordance with the Einstein equivalence principle, we have come to expect that all objects should fall identically under the influence of gravity, and this by no means excludes quantum particles exhibiting their wave-like nature. However, this does not discount the possibility of COW-like phases~\cite{Colella:1975jc} skewing the wavefunction at the screen to give \emph{apparent} violations.

This leads us to our first result, which is to explore how the phase generated by the gravitational potential results in an interference pattern that appears to fall like a classical object. It also provides the foundation for a more sophisticated problem explored in a later Section. From a conceptual point of view, this is an interesting scenario to investigate, especially when one considers the path integral formulation of quantum mechanics. 

In simple terms, the Feynman propagator is the Green's function for the Schr\"odinger equation, the solution resulting from the initial spatial wavefunction being a dirac-delta distribution. It represents the amplitude for a particle at position $x$ and time $t$ to be found at $cx\pri$ a later time $t\pri$. Once the propagator is known, the evolution for any initial wavefunction can be found by convolution with the propagator.
The one dimensional propagator is often expressed as
\begin{gather}
\bk{x\pri}{U(t\pri - t)}{x} = K_0(x\pri,t\pri;x,t) = \bigintsss\D \left(x(t)\right) \exp\left[\frac{i}{\hbar}\int_t^{t\pri} L\left(x(s),\dot{x}(s),s\right)ds\right],\tageq\label{eq:FeynmanPropDef}
\end{gather}
where $U(\delta t ) = \exp(-i\op{H}\delta t/\hbar)$ is the time evolution operator, $x(t)$ is a parametrised path in space, $\D\left(x(t)\right)$ is the Feynman measure over all possible paths and $L\left(x,\dot{x},t\right) $ is the Lagrangian describing the system. 

The path-integral formulation of quantum mechanics is conceptually very appealing, since it can be interpreted as a statement about how quantum mechanical objects may deviate from the laws of classical dynamics. In fact, even in the presence of a gravitational field, there is a non-zero amplitude which corresponds to the quantum system not falling at all: $\ip{x,t\pri}{x,t}>0 $ for some $t\pri>t$. Thus, from a foundational point of view, we would like to use the path integral approach to examine the way in which the gravitational potential affects a single-particle, double-slit interference pattern.

A short outline of the derivation is shown here, with full details available in Appendix~\ref{App:PathIntGravity}. The Lagrangian for a particle in a Newtonian gravitational potential is $L= \frac12 m \dot{x}^2 - m g x$. With reference to the propagator defined in Eq.~\eqref{eq:FeynmanPropDef}, we parametrise the path $x(t)$ in terms of deviations $\delta x(t)$ from the classical trajectory, $x_c(t)$, between the two points. This gives $x(t) = x_c(t) + \delta x(t)$, with $\delta x(t)=\delta x(t\pri) = 0$. This parametrisation leaves the Feynman measure unchanged, as a sum over all paths is equivalent to a sum over all deviations from a specific path. 
We are then left with two terms: a phase dependent on the action of the classical trajectory and a Feynman integral over the deviations that has a form identical to that of a free particle. We substitute the integral with the free particle propagator, but acknowledge that, since this is sum over deviations, we must set $x=x\pri=0$. 
The propagator for a particle in a Newtonian gravitation potential is then
\begin{gather}
K_g(x\pri,t\pri;x,t)= \frac{\exp\left[\frac{i}{\hbar} S\left[x_c(t)\right]\right]}{\sqrt{2\pi i \hbar (t\pri - t)/m}},\tageq\label{eq:GgProp}
\end{gather}
where $S\left[x_c(t)\right]$ is the functional that gives action associated with the classical trajectory between the points.
We can express it as a function of $(x,t,x\pri,t\pri)$ by solving the equations of motion for the boundary conditions $x_c(t)= x$ and $x_c(t\pri)= x\pri$. With complete details in Appendix~\ref{app:PropDeriv}, the general form for the classical action is given by,
\begin{gather}
S[x_c(t)] = \frac{m}{2}\left\{\frac{(x\pri - x)^2}{t\pri-t} - g(x+x\pri)(t\pri-t) - \frac{g^2}{12}(t\pri-t)^3\right\}.\tageq\label{eq:Sclassical}
\end{gather}

\subsection*{Single and double slit interference}

We now consider applying this propagator to the problem at hand. Let's begin by assuming that the slits are long enough to ignore diffraction effects in the $y$-direction (perpendicular to the gravitational field -- which is in the negative $x$-direction -- but in the plane of the screen), this allows us to effectively reduce the problem to two dimensions. Consider a source of particles at the origin $(0,0)$ and let a double slit be located at distance $D$ from the source in the $z$-direction. Each slit has width $2a$ with centre located at $x=\pm b$. The screen is then a further distance $L$ away from the slits. The two-dimensional propagator required for this problem is given by a free particle propagator in the $z$-direction, multiplied by the gravitational propagator for the $x$ direction, as calculated in Eq.~\eqref{eq:GgProp}. This propagator allows us to ask the question: \emph{If a particle initially starts at position $\vr = (0,0)$, what is the probability of finding it at position $\vrp=(x,D+L)$ on the screen?} This distribution in $x$ will be the the two slit interference pattern that we seek. 

When computing this amplitude, we consider a semi-classical approach. We assume that the `trajectory' of the neutron can be separated into two parts: (a) the path from the source to the slits, followed by (b) the path from the slits to the screen. Quantum mechanically, the particles need not pass through the slits and there even exists the possibility of them passing through the slits multiple times before hitting the screen. That being said, the probabilities associated with these events are negligible under certain conditions: The semi-classical approach is valid, provided that the majority of the particle's momentum is in the $z$ direction, such that the wavelength is approximately the $z$-direction wavelength, $\lambda \approx \frac{2\pi \hbar}{m v_z}$. We assume that this wavelength is much smaller than the relevant $z$-direction length scales, $D$ and $L$, in conjunction with the assumption that these are much larger than the relevant $x$ direction length scales, $b$ and $a$. Within this regime, the problem reduces to a single dimension.  After a rather tedious calculation (included in Appendix~\ref{sec:SlitWF} for completeness), the wavefunction at the screen due to a single slit centred at $x=b$, the instant the particle hits it in the semi-classical approximation ($\tau=L/v_z$), is given by
\begin{gather}
\psi^\text{(1)}(x) = \frac{e^{i\phi(x)}}{i2\sqrt{\eta a}} \bigg\{C[\sigma_+(x)] - C[\sigma_-(x)] + iS[\sigma_+(x)]-iS[\sigma_-(x)]\bigg\},\tageq\label{eq:psi1}
\end{gather}
where $C[u] \equiv \int_0^u \cos\left(\frac{\pi}{2}x^2\right)dx$ is the Fresnel cosine function, $S[u] \equiv \int_0^u \sin\left(\frac{\pi}{2}x^2\right)dx$ is the Fresnel sine function and $\eta = 1 + \frac{L}{D}$. Above
\begin{align}
&\sigma_{\pm}(x) = \sqrt{\frac{2}{\lambda L}\eta}\left\{(b\pm a) -\frac{x}{\eta} - \frac12g \frac{m^2 \lambda^2}{h^2}DL\right\} \tageq\label{eq:sigmapm}
\quad \quad \mbox{and}\n
&\phi(x) = \pi \left\{\frac{x^2}{\lambda(D+L)} - m g x\frac{\lambda(D+L)}{h^2} - \frac{g^2}{12}\frac{m^4\lambda^3}{h^4}(D+L)(D-L)^2 \right\}\tageq\label{eq:phix}.
\end{align}
If $b$ is set to zero, then this gives single slit diffraction. Extension to double slit or even $N$-slit interference is given by taking a normalised superposition of the wavefunctions corresponding to the different slit positions.

\begin{figure}[!htb]
\begin{center}
\includegraphics[width=0.96\textwidth]{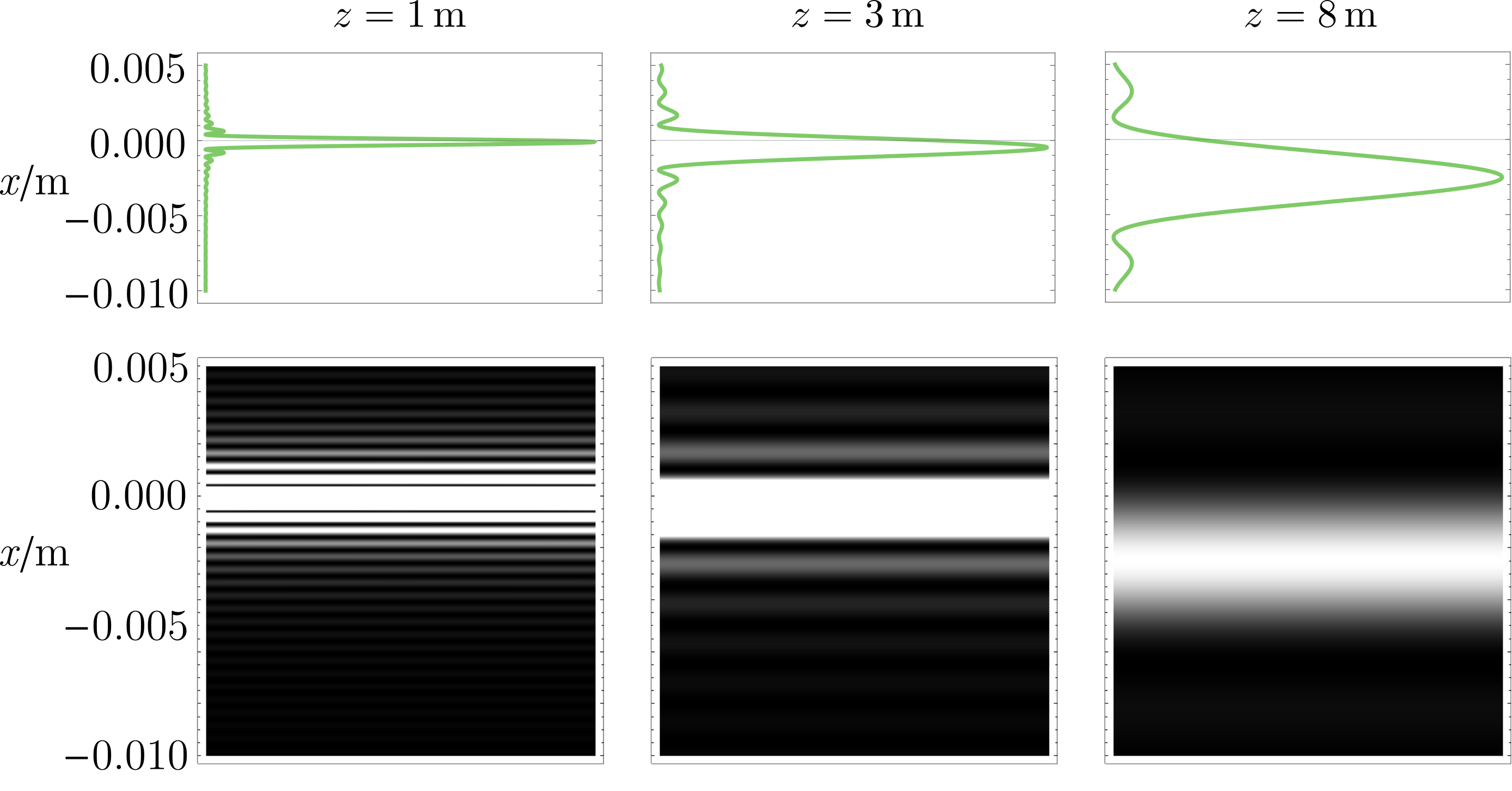}
\caption{\textbf{Single Slit Diffraction in a Gravitational field.} In the top row, the magnitude squared of the wavefunction in Eq.~\eqref{eq:psi1} $|\psi^{(1)}(x)|^2$ is plotted for source to screen distance $D=2\, {\rm m}$ and slit to screen distances $z=\{1\, {\rm m},3\, {\rm m},8\, {\rm m}\}$. The second row shows the same information as a two-dimensional probability density on the screen. The particle was chosen to be a neutron with wavelength $\lambda \sim 10^{-9}\, {\rm m}$, and the gravitational field strength $g=9.8\, {\rm m s}^{-2}$. In addition to the typical spreading of the pattern, we observe an apparent translation of the pattern, which we can interpret as falling. }
\label{fig:FallingDiff}
\end{center}
\end{figure}

The square of this wavefunction will give the observed probability distribution for the position at which the particle hits the screen; this is plotted for a single slit in Figure~\ref{fig:FallingDiff}. The pattern clearly appears to shift towards the negative $x$ direction as the screen is moved further from the slit. In general, this is far easier to identify in single slit diffraction, as the spreading of the pattern is less noticeable than in the double slit case.

The location of the central maximum is indicative of the position at which a classical point particle would arrive. If we consider a single particle incident on the slit, then there exists a possibility that it will be detected above the central maximum. We could interpret this as the particle having fallen less than expected classically. Similarly we could detect a particle below the central maximum, indicating that it fell faster than expected classically. Although it would be appealing to label this as violation of the equivalence principle, to do so would be incorrect. The easiest way to verify this is to transform to an accelerated coordinate system, taking us to the freely falling frame, in which gravitational effects should completely vanish.

When calculating the propagator in Eq.~\ref{eq:GgProp}, we made use of the general form for the classical trajectory, which satisfies the Euler-Lagrange equations of motion. Using the solution, we find that the classical parabolic trajectory an object takes from the source at $(x=0,t=0)$ to the slit at $(x=0,t=T)$ requires an initial upward speed $v_x(t=0) =  \frac12 g T$ and final vertical speed of $v_x(t=T)= - \frac12gT$. The position of the object $\tau$ seconds later would then be,
 \begin{gather}
x_c(T+\tau) = -\frac{g\tau}{2}g(\tau+T)
= -\frac{g m^2\lambda^2 }{2h^2}\left(L^2 + LD\right). \tageq\label{eq:ClassicalDisp}
\end{gather}
We would expect that, by performing the coordinate transformation $x = \xi + x_c(T+\tau)$, the pattern should become identical to the case where $g=0$, in accordance with the equivalence principle. Since the $x$ dependence in the wavefunction $\psi^\text{(1)}$ appears only through the function $\sigma_{\pm}(x)$, we can work directly with the expression given in Eq.~\eqref{eq:sigmapm},
\begin{align}
\sigma_{\pm}(\xi) &= \sqrt{\frac{2}{\lambda L}\eta}\left\{(b\pm a) -\frac{\xi -  \frac12g\frac{m^2\lambda^2 D^2 }{h^2}\left(L^2 + LD\right)}{\eta} - \frac12g \frac{m^2 \lambda^2}{h^2}DL\right\}\n
&= \sqrt{\frac{2}{\lambda L}\eta}\left\{(b\pm a) -\frac{\xi}{\eta} + \frac12g\frac{m^2\lambda^2 D^2 }{h^2}L\left[\frac{D\left(L + D\right)}{L+D}-D\right]\right\}\n
&= \sqrt{\frac{2}{\lambda L}\eta}\left\{(b\pm a) -\frac{\xi}{\eta} \right\}.\tageq\label{eq:sigmaXI}
\end{align}
The result above shows that the gravitational effect on the interference can be eliminated by transforming to an accelerated coordinate system. It is now clear that there are no equivalence principle violations; if we detect a particle away from the central maximum, it is interpreted as the usual deviations of a quantum particle from its classically expected trajectory.

We don't need to calculate the double slit pattern to identify the absence of a COW phase. Since the multi-slit wavefunction is simply a superposition of single slit wavefunctions, the coordinate transform above extends to the general case;  the only effect gravity will have is a translation of the entire pattern. The reason for this is that in the COW experiment, the Mach-Zehnder interferometer constrains the path of the particle to an approximate binary. Whilst confined to these paths, the relative Aharonov-Bohm-like phase is accumulated. In the case of single-slit diffraction, there is no path confinement, and even for multiple slits, where there are discrete variations between paths, there is no relative phase accumulated; this is because the slits are effectively infinitely thin in our scenario. 

Therefore, it would appear that there are no peculiar quantum effects that appear in a freely falling interference pattern, beyond what one would expect in the absence of gravity. The quantum mechanical deviations from the classically expected trajectory represent a departure from the laws of classical physics, and, although the deviations might seem to constitute a violation of the universality of free fall, the effects are completely consistent with quantum behaviour as viewed from an accelerating coordinate system. In other words, an interference pattern falls like a classical object. 

\section*{Effects of internal degrees of freedom}

In this Section, we examine some of the gravitational effects that appear at leading relativistic order for particles with internal degrees of freedom. These effects, which can be seen as arising from relative time dilation of different internal levels, were first investigated in detail by Zych \textit{et al.}~\cite{Zych:2011jz,Zych:2012kq} and Pikovski \textit{et al.}~\cite{Pikovski:2015du}, and were further discussed by Zych and Brukner in the context of the equivalence principle~\cite{Zych:2015vm}.

\subsection*{The Hamiltonian formulation}

According to the Einstein equivalence principle, all internal energy acts as a mass from the perspective of both general and special relativity. That is, the mass terms appearing in the kinetic and potential energy of a system in a gravitational field should depend on the internal energy state. When the internal state corresponds to a dynamically varying degree of freedom, with its own Hamiltonian $H^\text{int}$, then all terms involving the mass should couple it to the external degree of freedom. In other words, the mass is promoted to an operator on the internal degree of freedom:
\begin{gather}
m\rightarrow M = m \, \id^\text{int} + \frac{H^\text{int}}{c^2}.
\end{gather}
The full Hamiltonian for a particle in a uniform gravitational field, including the newly defined mass operator is then (to leading relativistic order)~\cite{Zych:2015vm}
\begin{align}
H &= M c^2+ \frac{P^2}{2M} + M  g x\n
&= \left(m\id^\text{int} + \frac{H^\text{int}}{c^2}\right)c^2+ \frac{P^2}{2\left(m \id^\text{int} + \frac{H^\text{int}}{c^2}\right)} + \left(m \id^\text{int} + \frac{H^\text{int}}{c^2}\right) g x\n &= m c^2 + H^\text{int} +  \frac{P^2}{2m } + m  g x + \frac{1}{m c^2}\left\{-\frac{P^2}{2m}H^\text{int} + g x H^\text{int} \right\}  + \mathcal{O}\left(c^{-4}\right),\tageq\label{FOHam}
\end{align}
where, in the last line, we have expanded the $M^{-1}$ in a Taylor expansion. This is valid, provided that the largest eigenvalue of the internal Hamiltonian, denoted by $\|\hint\|$, satisfies $\|\hint \|/m c^2\ll 1$, \textit{i.e.}, the internal energy is small compared to the rest mass. The additional terms introduced by the mass operator give lowest order relativistic effects. The first effect is introduced by the coupling of the internal energy to the kinetic energy operator, which represents lowest order special relativistic time dilation. The other interaction term, coupling the internal energy to the Newtonian potential, represents lowest order gravitational time dilation effects. 

We can verify this by looking at the evolution of the internal degree of freedom. Provided that the internal evolution is not trivial, i.e., that it is not in an eigenstate of the internal Hamiltonian, it can be considered operationally as a clock~\cite{Pikovski:2015du}. If we denote $q$ to be an observable of the internal degree of freedom, then the evolution given in the Heisenberg picture is, as described in Ref.~\cite{Zych:2015vm},
\begin{align}
\dot{q} = \frac{1}{i\hbar}[q,H] &= \frac{1}{i\hbar}\left\{[q,\hint]\id^\text{ext} - [q,\hint]\frac{P^2}{2m^2c^2} + [q,\hint]\frac{gx}{c^2}\right\}\n
&= \dot{q}_\text{loc}\left(1 - \frac{P^2}{2m^2c^2} + \frac{g \,x}{c^2}\right).\tageq\label{eq:internalEvol}
\end{align}
Here $\dot{q}_\text{loc}$, is the normal rate of internal evolution as given in the system's rest frame. Recalling that the rate of change of proper time, in the non-relativistic, weak-field limit, is $d\tau = (1 -\frac{v^2}{2c^2} - \frac{\phi(x)}{c^2})dt$, we can easily identify these additional terms as a result of lowest order time dilation. For semi-classical evolution of the external degrees of freedom, the evolution of the internal degree of freedom is affected in a manner that is consistent with our understanding of relativistic effects. Interestingly this equation is valid not just for semi-classical systems, but also for non-local systems or systems with momentum that is not well defined. In these cases however, we cannot apply any of our classical intuition~\cite{Pikovski:2015du}.

This result can be interpreted in the following way; general relativity provides a description for the evolution of clocks which are attached to observers evolving according to the laws of classical mechanics. On the other hand, the mass operator, has in a sense, allowed us to describe the evolution of a clock attached to an observer who evolves according to the laws of quantum mechanics.

Though this intuition can be applied to the internal evolution, we will present results that show this is not true when observing the external evolution. The evolution of the position degree of freedom is given by,
\begin{gather}
\dot{x} = \frac{1}{i \hbar} [x,H] = \frac{1}{i\hbar}\frac{[x,P^2]}{2m\left(1 + \frac{\hint}{mc^2}\right)}
= \frac{P}{m}\left(1 - \frac{\hint}{mc^2}\right) + \O\left(m^{-2}c^{-4}\right).\tageq\label{eq:Vel}
\end{gather}
Again, if we consider a semi-classical wavepacket, and take the expectation value of the equation above, we find that the velocity of this wavepacket depends on the state of the internal degree of freedom. In particular, a particle in an excited state will have a slower expected velocity than one in its ground state. If the particle is prepared in a superposition state of internal energy, then its position at a later time will be entangled with the internal degree of freedom. Thus, the mass operator introduces spatial decoherence, which even appears in the case of a free particle~\cite{Pikovski:2015du}.

\subsection*{The path integral formulation}

To examine these effects further, we will investigate the mass operator from the perspective of the path integral formalism. We motivate the work here with the question: \emph{How does the mass operator affect the falling interference presented in the last Section?} In order for there to be any effect, the particle must have some non-degenerate internal energy levels. We will restrict ourselves to the simple case discussed in Ref.~\cite{PJORLANDO2015}, where the particle is spin-$\frac{1}{2}$ with a Zeeman splitting induced by an external magnetic field\footnote{We will, however, still consider the particle to be neutral, so there is no coupling to the electromagnetic field beyond its spin interaction.}. Before we can answer the above question, we need to find the form for the propagator in this scenario. 

We begin with the newly defined Lagrangian for this problem,
\begin{gather}
L(x,\dot{x}) = \frac12M\dot{x}^2  - Mg x - Mc^2, \tageq\label{eq:LMOP}
\end{gather}
For a particle with magnetic moment $\mu$ in a uniform magnetic field of strength $B$, the mass operator is given by
\begin{gather}
M =  \left[\begin{array}{cc} m - \frac{\mu B}{2c^2} & 0 \\ 0 & m + \frac{\mu B}{2c^2}\end{array}\right].\tageq\label{eq:MOP}
\end{gather}
From this point, we can construct the Feynman propagator, in accordance with Eq.~\eqref{eq:FeynmanPropDef}. 
\begin{gather}
K^{\chi\pri\!,\chi}(x\pri,t\pri;x,t) = \bk{x\pri,\chi\pri}{U(t\pri - t)}{x,\chi}\tageq\label{eq:SpinProp}
\end{gather}
This is still a matrix element of the time-evolution operator; however, the evolution operator now contains an index for spin, accounting for the two-dimensional internal Hilbert space. This also naturally leads to a matrix representation:
\begin{gather}
K^{\chi\pri\!,\chi}(x\pri,t\pri;x,t) = \left[\begin{array}{cc}K^{\uarr\uarr}(x\pri,t\pri;x,t) & K^{\uarr\darr}(x\pri,t\pri;x,t) \n K^{\darr\uarr}(x\pri,t\pri;x,t) & K^{\darr\darr}(x\pri,t\pri;x,t) \end{array}\right]. \tageq\label{eq:DefMatrixProp}
\end{gather}
Since the interaction terms appearing in Eq.~\ref{eq:LMOP} all commute with the mass operator in Eq.~\ref{eq:MOP}, the propagator can be greatly simplified, as it is then diagonal in the internal energy eigenbasis:
\begin{gather}
\K(x\pri,t\pri;x,t) = \left[\begin{array}{cc}K^{m_-}(x\pri,t\pri;x,t) &0 \n 0& K^{m_+}(x\pri,t\pri;x,t)\end{array}\right],\tageq\label{eq:DefMatrixProp2}
\end{gather}
where $K^{m_\pm}(x\pri,t\pri;x,t)$ is the propagator for a particle of mass $m_\pm=m \pm \mu B/(2c^2)$; from the perspective of the propagator, the different internal energy states just appear as modified masses. Typically, the rest mass energy is excluded from the Lagrangian; it has no effect on the dynamics, and merely leads to an unmeasurable global phase $\exp(-{i mc^2 t}/{\hbar})$. When promoting mass to an operator, a relative phase of $\exp({-i \mu B t }/{\hbar})$ is introduced between the two internal states, which could in principle have a measurable effect.

We wish to use this propagator to calculate the wavefunctions for particles which initially have spin in the superposition $\ket{\chi_{_0}} = \alpha\ket{\uarr} + \beta\ket{\darr}$, with $|\alpha|^2 + |\beta|^2 =1$. However, if we are only interested in the interference pattern observed on the screen, and do not measure the spin state of the particle, then we need to trace out the spin information. We review how to do this in Appendix~\ref{app:PropagatorTrace}. We find that the spatial probability distribution for an initial state $\ket{\chi_{_0}}\otimes\ket{\psi_0}$ is then
\begin{gather}
\exx{x} = |\alpha|^2\left|\int dx K^{m_-}(x,t;x,0)\psi_0(x)\right|^2 +|\beta|^2\left|\int dx K^{m_+}(x,t;x,0)\psi_0(x)\right|^2, \tageq\label{eq:ConvexSum}
\end{gather}
which is a convex sum of the contributions coming from each internal state. This is immediately identifiable as decoherence, which is consistent with our interpretation of Eq.~\eqref{eq:Vel}. Additionally, this demonstrates that, when tracing out the spin degree of freedom, the phase introduced by the rest mass operator becomes irrelevant.

The form of Eq.~\eqref{eq:ConvexSum} allows for easy calculation of the decohered spatial distribution, which we will illustrate with an example. Take the propagator to be that of a free particle and choose the initial wavefunction to be a Gaussian wavepacket with momentum $p$. This wavefunction is given by
\begin{gather}
\psi_0(x) = {\left(\pi \sigma\right)^{-\frac14}} \exp\left[{-\frac{x^2}{2\sigma^2}+\frac{ip x}{\hbar}}\right].\tageq\label{eq:initialWF}
\end{gather}
After convolving this with the free space propagator $K_0^{m_\pm}(x,t,x_0,0)$, we have
\begin{gather}
\psi_{m_\pm}(x,t) = \frac{\exp\left[i\phi-2\frac{(z-p/m_\pm t)^2}{\sigma^2(1+\gamma_{m_\pm}^2)}\right]}{(\pi \sigma^2)^{1/4}\sqrt{i - \gamma_{m_\pm}}},
\end{gather}
where $\gamma_{m_\pm} = \frac{\hbar t}{m_\pm\sigma^2}$ and $\phi$ is an irrelevant phase factor. We notice that the mean of this Gaussian moves with speed $p/m_\pm$. If the particle is prepared in a superposition state of internal energy then Eq.~\eqref{eq:ConvexSum} states that the probability distribution will be given by
\begin{gather}
P(x,t)= |\alpha|^2\big|\psi_{m_-}(x,t)\big|^2 + |\beta|^2\big|\psi_{m_+}(x,t)\big|^2.
\end{gather}
In other words, the spatial distribution is given by a mixture of Gaussian wavepackets propagating with different speeds. 

Given the initial state $\ket{\Psi} = (\alpha\ket{\uarr} + \beta\ket{\darr}) \otimes \ket{\psi_0}$, the coupling introduced by the mass operator evolves the state to $\ket{\Psi(t)} = \alpha\ket{\uarr}\otimes\ket{\psi_{m_-}(t)} + \beta\ket{\darr} \otimes \ket{\psi_{m_+}(t)}$. If a detector is placed at a distance from the source far enough for the Gaussian distributions described by $\ip{x}{\psi_{m_-}(t)}$ and $\ip{x}{\psi_{m_+}(t)}$ to become distinct, then the arrival time of the particle will be bimodal. Again, if the position degree of freedom is considered to be a `clock' -- its non-trivial evolution permits this -- then this may be considered to be a special relativistic time dilation effect~\cite{Pikovski:2015du}.

\section*{Gravitational decoherence in double slit interference} 

We have now developed the tools to explore falling double slit interference with an internal degree of freedom. Again, we consider a particle incident on slits of width $2a$ centred at $x=\pm b$. Our earlier calculation in the first Section used the semi-classical approximation for the $z$-direction, to replace arrival times $T$ and $\tau$ with the classically expected times, $D/v$ and $L/v$. We have just shown, however, that the arrival time of the particle will no longer be well defined when the internal degree of freedom plays a dynamical role. 

We also saw, in the previous Section, that the effect of using the gravitational propagator was equivalent to performing a mass-independent coordinate transformation. This means that, for a wavepacket with zero initial average momentum, the mass operator has no effect on the position expectation value under the influence of a gravitational potential\footnote{It will however affect the spreading of the wavepacket and therefore the variance in the position.}. This leads to an interesting effect if we consider a two dimensional Gaussian wavepacket with zero average momentum in the $x$-direction (in the direction of the gravitational field) and a non-zero average momentum in the $z$-direction. At some fixed distance along $z$ from the particle's initial location, the difference in expected arrival times will mean that, depending on the internal state, gravity will have displaced the wavepacket for different amounts of time. As a result, the higher energy state will fall further than the lower energy state, causing gravity to act, in some sense, like an asymmetric Stern-Gerlach device. This effect will be very small, as it depends on the magnitude of $\|\hint\|/(mc^2)$, but can be sensitively detected by introducing an interference pattern along $x$.

\begin{figure}[t]
\begin{center}
\includegraphics[width=0.96\textwidth]{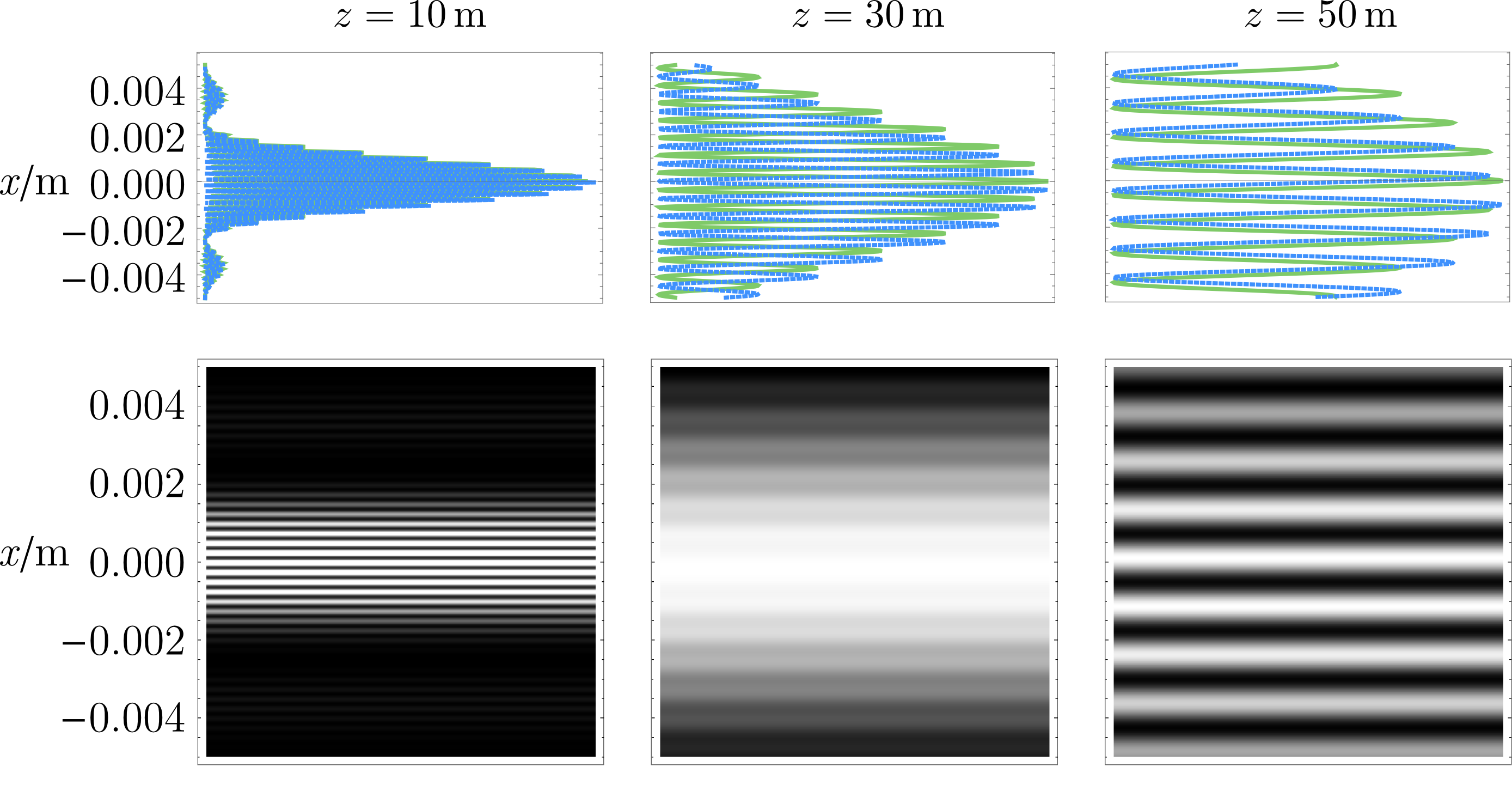}
\caption{\textbf{Decoherence of Double Slit Interference in a Gravitational Field.} The internal energy splitting leads to different expected arrival times for the  wavepacket at the screen. The pattern corresponding to the higher energy spin state will fall further than its counterpart, resulting in periodic reductions in the visibility of the interference. The intensity at the screen for distances of $10\,{\rm m}$, $30\,{\rm m}$ and $50\,{\rm m}$ is shown for a neutron with wavelength $\lambda \sim 10^{-8}\,{\rm m}$ and internal energy splitting of $\Delta E = 10^{-14}\,{\rm J}$. The top row shows the time-averaged spatial probability distribution for the up (green, solid) and down (blue, dotted) spin components, while the bottom row plots the spin-averaged probability distribution as it would be observed on the screen. All $x$ positions are relative to the position of a classical particle with mass $m_-$.}
\label{fig:IntDec}
\end{center}
\end{figure}

We calculate the pattern produced by the double slit by returning to a two dimensional propagator, and simplifying the problem to a Gaussian particle distribution incident on the slits which is then detected at a screen $L$ metres away. In this case, the wavefunction just beyond the slits, for a particle of mass $m$, is given by
\begin{gather}
\psi^\text{(1)}\left(x,z,t\right) = \frac{\int_{b-a}^{b+a}\int_\ninfty^\infty dx_0 dz_0 K_0(z,t;z_0,t)K_g(x,t;x_0,0)\psi_0(x_0,z_0)}{\int_{b-a}^{b+a}dx \psi(x_0,0)}.\tageq\label{eq:Gauss2DDiff}
\end{gather}
This avoids the semi-classical approximations made in the previous calculation, but leads to a time-dependent wavefunction. However, we are only interested in the spatial distribution observed at the screen, with no measurement performed regarding the time of arrival. The simplest way to account for this is to average the distribution over some length of time so that
\begin{gather}
\bar{\psi}(x,z=L) = \frac{1}{2\Delta t}\int_{-\Delta t}^{\Delta t} |\psi(x,L,t)|^2dt,\tageq\label{eq:TimeAveraged}
\end{gather}
where $\Delta t$ will have some relationship with the spatial spread of the Gaussian packet in the $z$ direction, such that the majority of the probability lies within $\pm\Delta t$. Figure~\ref{fig:IntDec} shows a plot of the resulting two-slit interference, calculated for a neutron in equal superposition of its internal energy states. The energy splitting is $\Delta E \approx 10^{-14}\,{\rm J}$, corresponding to a magnetic field on the order of $10^{12}\, {\rm T}$. Even with this infeasibly large energy splitting, the decoherence effect occurs over tens of metres. If a more reasonable value for the energy splitting is used, then spreading of the wavepacket delocalises the particle before the decoherence is even detectable.

This decoherence is suggestive of the effects described in Ref.~\cite{Zych:2011jz}; where Zych \emph{et al.} demonstrate that the interference pattern in a Mach-Zehnder interferometer decoheres as a result of proper time. The work we present here is a free space interference analogue, which similarly exhibits periodic decoherence effects. This gives strength to the argument in Ref.~\cite{Pikovski:2015du} that the effects on external evolution, introduced by the mass operator, are complementary to the interpretation of time dilation, occurring for evolution of the internal degrees of freedom. To put it more elegantly, embedding an operational clock in a system which behaves quantum mechanically results in an evolution which destroys this quantum nature.

\section*{Coherence, correlations, and non-Markovian dynamics}

The decoherence of the interference fringes discussed above can be better understood in terms of correlations between internal and external degrees of freedom. It also turns out that spin coherence is necessary for the generation of non-classical correlations between the internal and external degrees of freedom. We discuss each of these ideas successively below, beginning with coherence theory, before providing a further interpretation in terms of non-Markovian open dynamics.

\emph{Incoherent operations.} The total Hamiltonian in Eq.~\eqref{FOHam} is diagonal in the spin basis and can therefore be expressed as $H = \kb{\uarr}{\uarr} \otimes H_{m_{-}} + \kb{\darr}{\darr} \otimes H_{m_{+}}.$ Therefore, the unitary operator for the joint dynamics has the form of a controlled-unitary on the external degree of freedom:
\begin{gather}\label{eq:incoherent-U}
U = \kb{\uarr}{\uarr} \otimes U_{m_{-}}
+ \kb{\darr}{\darr} \otimes U_{m_{+}},
\end{gather}
which has the potential to generate correlations between the internal and external degrees of freedom. However, when considering the reduced dynamics of the spin alone, evolution between two points in time is described by an incoherent operation (one which maps incoherent states to incoherent states) in the energy eigenbasis~\cite{arXiv:1609.02439}, \textit{i.e.}, 
\begin{gather}
U\ket{s\, \psi(0)} = \ket{s\, \psi_{m_r}(t)},
\end{gather}
where $s\in\{\darr,\uarr\}$, $r=+$ if $s=\darr$,  and $r=-$ if $s=\uarr$. More specifically, $U$ is incoherent since it will map a mixed state of the form $\sigma_{IC}=q \kb{\uarr}{\uarr} + (1-q) \kb{\darr}{\darr}$ to itself: 
\begin{gather}
\tr_{ext}[U \sigma_{IC} \otimes \rho U^\dag] = \sigma_{IC}, 
\end{gather}
where $\rho$ is any state of the external degree and $IC$ stands for incoherent.

On the other hand, due to the non-vanishing commutator $[P^2,x] \ne 0$, the dynamics of the external degree of freedom alone is not described by an incoherent operation in the position basis. That is, 
$U\ket{s\, x} = \ket{s\, \psi_{m_r}(t)}$,
and coherence of the wavefunction can increase.   For the same reason, both conditional unitary operations $U_{m_{-}}$ and $U_{m_{+}}$ will also look like incoherent operations from the perspective of the spin.

\emph{Entanglement and discord.} In recent years, researchers studying coherence theory have shown that incoherent operations can lead to generation of entanglement and quantum discord when the initial spin state possesses coherence. The generation of entanglement is easily checked by taking the spin to initially be in the pure state $\alpha \ket{\darr} + \beta \ket{\uarr}$ for $\alpha,\beta \ne 0$ and the position state to be $\ket{\psi}$:
\begin{gather}
U\left( \alpha \ket{\darr \, \psi(0)} + \beta \ket{\uarr\, \psi(0)}\right) = \alpha \ket{\darr \, \psi_{m_{+}}(t)} + \beta \ket{\uarr\, \psi_{m_{-}}(t)},
\end{gather}
which is an entangled state, since the marginal states are not pure. 

It is known that any coherence can be turned into entanglement via some incoherent operation and a pure ancilla~\cite{PhysRevLett.115.020403}. However, in our setup we are limited to a specific incoherent operation, which may not be able to generate entanglement for all coherent initial states. Consider the case when the initial spin state is the mixed state 
\begin{gather}
\sigma_{CO} = w \kb{\uarr}{\uarr} + (1-w) \kb{+}{+},
\end{gather}
where  $\ket{+}=(\ket{\uarr}+\ket{\darr})/\sqrt{2}$ and $CO$ stands for coherent. For $0 \le w \le 1$, the time-evolved state will have quantum correlations, but for some values of $w$ will have no entanglement; the future state will be fully separable for $w=1$ and entangled for $w=0$. Therefore, there must be a critical value for $w=w_c$ where the transition from entangled state to separable state occurs. In the regime where the state is separable, it will necessarily have quantum discord~\cite{arXiv:1107.3428, RevModPhys.84.1655, arXiv:1605.00806} as measured by the internal or external degree of freedom.

In fact, the only time quantum discord vanishes for $t>0$ is when the initial spin state has the form $\sigma_{IC} = q \kb{\uarr}{\uarr} + (1-q) \kb{\darr}{\darr}$. Let us further suppose that the initial external state is given by a density matrix $\rho$. After evolving for some time $t$, the system will be in state
\begin{gather}\label{eq:clstate}
U\sigma_{IC} \otimes \rho \, U^\dag =
q \kb{\uarr}{\uarr} \otimes \rho_{m_-}(t)+ (1-q) \kb{\darr}{\darr} \otimes \rho_{m_+}(t).
\end{gather}
This clearly becomes a classical mixture of the states $\rho_{m_-}(t)$ and $\rho_{m_+}(t)$ with weighting $w$ when the internal degree of freedom is measured (whichever measurement basis is chosen). That is, the unitary operation in Eq.~\eqref{eq:incoherent-U}, being an incoherent operation on the internal degree of freedom, will not generate any non-classical correlations when the initial spin state is a classical mixture of energy eigenstates. 

On the other hand, with the exception of pathological cases where the initial wavefunction does not have support everywhere, the two spin-conditioned external states will never be exactly orthogonal, \emph{i.e.}, $\tr[\rho_{m_+}(t)\rho_{m_-}(t)]\ne 0$. This means that the spin state after a measurement on the external degree of freedom will, in general, depend on the choice of measurement basis; in other words, there are non-classical correlations (discord) in one direction.

Let us now consider the case where both initial states can be arbitrary mixed states. Then the time evolved states have form
\begin{gather}\label{eq:totalstate}
U 
\begin{pmatrix} a & b \\
b^* & 1-a \end{pmatrix} 
\otimes \rho \, U^\dag = 
\begin{pmatrix} a \, \rho_{m_-} & b\, U_{m_-} \, \rho \, U_{m_+}^\dag\\
b^* U_{m_+} \, \rho \, U_{m_-}^\dag & (1-a) \rho_{m_+}.
\end{pmatrix} 
\end{gather}
If we can make (non-unitary) operations on the spin degree of freedom, such as projections, we will see different interference patterns corresponding to different outcomes. By making strong measurements on the spin by, \emph{e.g.}, introducing a Stern-Gerlach apparatus, the correlations could be used to steer the interference pattern. 

For example, to see how correlations affect the reduced dynamics, consider the middle column of Figure~\ref{fig:IntDec}. If the magnetic field, and hence the effective coupling, was turned off for $z>30m$, the interference pattern would subsequently evolve unitarily with a single mass-$m$ propagator. It would continue to fall as if it were a classical object, and the fringe visibility would never return. This is because the spin degree of freedom, which `remembers' the original two-slit pattern, is no longer interacting with the position degree of freedom. However, if the spin components were filtered out at a later time using a Stern-Gerlach apparatus, the visibility could be recovered in full; the spin acts as a memory, hiding information about the particle's earlier trajectory. In other words, the periodic re-coherence of the spatial wavefunction is indicative of non-Markovian behaviour, which we will now discuss further.

\emph{Reduced non-Markovian dynamics.} In order to see the effects of these correlations from another perspective, suppose we only look at the position of the particle.
From the perspective of an observer who cannot measure the internal degree of freedom, the evolution of the particle appears to be open, with the spin acting as an environment. We immediately see that the same features are seen whether the initial spin state possesses coherence or not. The external state is obtained by tracing over the spin in Eq.~\eqref{eq:totalstate} to get
\begin{gather}\label{eq:marginalstate}
\rho_{ext} = a \rho_{m_-} + (1-a) \rho_{m_+}.
\end{gather}
That is, the observed interference pattern is indistinguishable and independent of $b$. When $b=0$ the internal and external degrees of freedom become classically correlated, and both entanglement and discord are vanishing. Conversely, when $b \ne 0$ discord (and possibly entanglement) will be present.  

Whenever the past state of a system directly affects its future evolution, a process is called non-Markovian. While there have been several mathematical definitions of `non-Markovianity' proposed for quantum processes~\cite{NMRev,breuer-rev} (with variable levels of descriptive success), the operational meaning of the term is clear cut~\cite{arXiv:1512.00589}: If the causal continuity of a system's evolution is broken at some time $t$ by, for example, making a measurement and re-preparing the system in a fixed state $\ket{\phi}$, independently of the measurement outcome, then the process is non-Markovian if the system's density operator $\rho_\tau(x,x\pri)$ at a later time $\tau$  depends on the measurement outcome $k$ or on the system's history $h$ prior to $t$. Formally,
\begin{gather}
    \rho_\tau(x,x\pri\,|\ket{\phi},k,h)\neq\rho_\tau(x,x\pri\,|\ket{\phi},k\pri,h\pri) \Rightarrow\; \text{Non-Markovian}.
\end{gather}
This kind of behaviour implies that there is some sort of memory transmitting information from the past across the causal break. We will see that this is the case for the falling particle described earlier in this Section.

\begin{figure}[t]
\begin{center}
\includegraphics[width=1\textwidth]{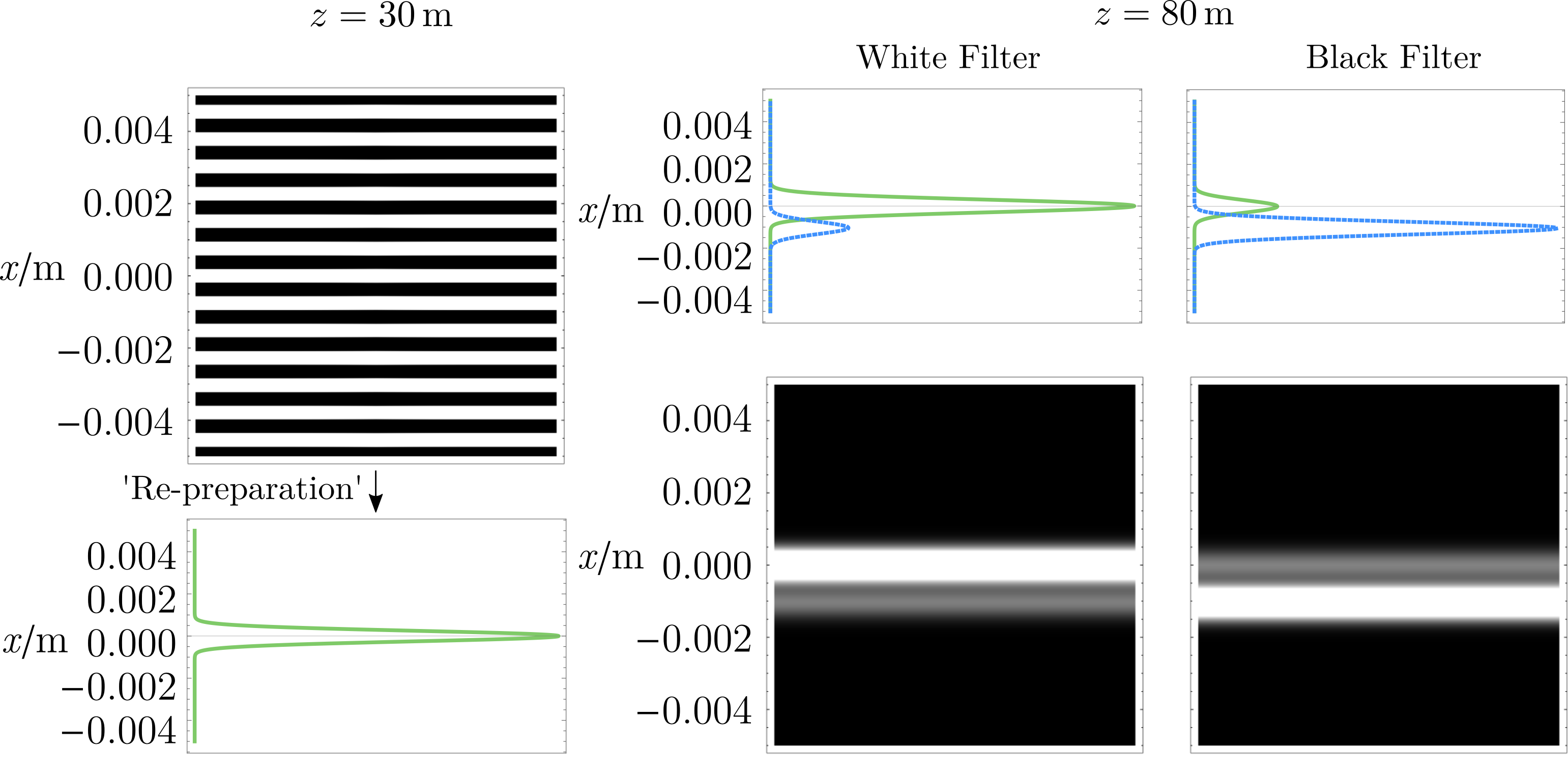}
\caption{\textbf{Non-Markovian Behaviour of an Interfering Particle with Spin.} A grating filter is introduced at $z=30\,{\rm m}$, allowing particles to pass through either the white or black regions in the top-left panel. The particle is then subsequently re-prepared in a Gaussian state, whose probability density is shown in the bottom-left panel; this state is the same for either choice of filter. The plots on the right show the probability distribution to find the particle at different positions on a screen placed at $z=80\,{\rm m}$ for the two choices of filter; top curves show spin-up (green, solid) and spin-down (blue, dotted) projections, the lower plots show the spin-averaged position distribution. Distinguishability of the two cases indicates non-Markovian behaviour. All $x$ positions are relative to the corresponding position of a classical particle with mass $m_-$.}
\label{fig:CausalBreak}
\end{center}
\end{figure}

In order to introduce a causal break in the evolution, we will put the spatial filter shown in the left column of Figure~\ref{fig:CausalBreak} at $z=30m$. This can be set to either allow particles through the white region (which has the greatest overlap with the spin-up wavefunction) or the black region (which has the greatest overlap with the spin-down wavefunction). After the filter, the particle is rapidly (effectively instantaneously) collimated into a Gaussian state along $x$, which does not depend on whether the black or white filter is chosen.

Since the overlap of the white (black) filter function $f_{w(b)}(x)$ with the spin-up and spin-down wavefunctions at $z=30m$ is different, the subsequent spin state will be conditioned on the choice of filter. The post-filter spin density operator is given by
\begin{align}
\rho_{w(b)} =& |\alpha|^2\left|\int {\rm d} x\,f_{w(b)}(x)\psi_{m_-}(x)\right| \kb{\uparrow}{\uparrow} + |\beta|^2\left|\int {\rm d} x\pri\,f_{w(b)}(x)\psi_{m_+}(x)\right| \kb{\downarrow}{\downarrow} \nonumber \\
&+\alpha\beta^*\int {\rm d} x {\rm d} x\pri\,f_{w(b)}(x)\psi_{m_-}(x)f_{w(b)}(x\pri)\psi_{m_+}^*(x\pri)\kb{\uparrow}{\downarrow}+ {\rm h.c.},
\end{align}
where $\psi_{m_\pm}(x)$ is the time-averaged wavefunction for the relevant spin branch at the $z$ position the filter is applied. For $\alpha=\beta=1/\sqrt{2}$, the post-filter probabilities for the spin-up and down states are $\sim \frac{4}{5}$ and $\sim\frac{1}{5}$ respectively for the white filter, and \emph{vice versa} for the black filter.

The right hand side of Figure~\ref{fig:CausalBreak} shows the probability distribution further from the slits after each of the filters is applied (note that the two spin components have already begun to separate again). Since the two conditional distributions are clearly different, the dynamics of the spatial distribution must be non-Markovian; the only way the post re-preparation evolution can depend on which filter was applied is through the spin state, which is acting as a memory.

\section*{Conclusion}

The universality of free fall is a pervasive phenomenon, and one which has inspired more fundamental gravitational equivalence principles. This includes Einstein's famous equivalence between mass and energy which, ultimately, forms part of the foundation for our current understanding of gravity. Here, we have explored how a self-interfering quantum particle falls under the influence of Newtonian gravity. We have shown that the universality of free fall holds even in this case, as the interference pattern itself fall just like a classical particle.

We have also considered interference of falling neutrons in the presence of a strong magnetic field. The presence of the magnetic field leads to splitting of the internal energy of the neutrons which, according to the Einstein equivalence principle, should make spin-down neutrons more massive than spin-up neutrons. Moreover, if a neutron is prepared in a spin-superposition state (with respect to the internal energy eigenbasis), this seemingly leads to the violation of a super-selection rule, \emph{i.e.}, superposition of masses. However, we use the mass operator formalism~\cite{Zych:2015vm} to show that, if the energy splitting of the internal spin contributes to the mass of the neutron, then the visibility of the interference pattern periodically decreases and increases.

Our results indicate that these decoherence effects are a consequence of an operational clock embedded within a quantum mechanical rest frame. That is, the internal degree of freedom keeps track of the time the particle spends being in different mass states. Finally, we have shown that this accounting of the internal energy (mass) state can be understood as non-Markovian dynamics for the position degree of freedom, with the spin acting as a memory. We show the non-Markovian behaviour by operational methods using the notion of causal break introduced in Ref.~\cite{arXiv:1512.00589}. In particular, we have given an operational recipe to witness the non-Markovian memory by solely acting on the external degree of freedom.

In Ref.~\cite{Pikovski:2015du} it is argued that gravity may be the culprit for quantum decoherence. This mechanism does not depart from how we think of decoherence in open systems theory more generally. This view is fundamentally different from that posited by the proponents of collapse theory who claim that gravity leads to fundamentally irreversible dynamics, \emph{cf.} Ref.~\cite{Snadden:1998bj}. Thankfully, one can differentiate between the two hypotheses by checking whether the decoherence can be reversed~\cite{arenz_distinguishing_2015}, which we do here demonstrating that coherence-information loss due to gravity can be recovered.

\section*{Acknowledgements}
We thank Lucas Celeri and Robert Mann for insightful discussions, German Valencia for pointing out an error in an earlier version of this work.

\appendix

\section*{Appendices}

\section{The COW Experiment} \label{sec:COW}

\begin{figure}[t]
\begin{center}
\includegraphics[scale=0.7]{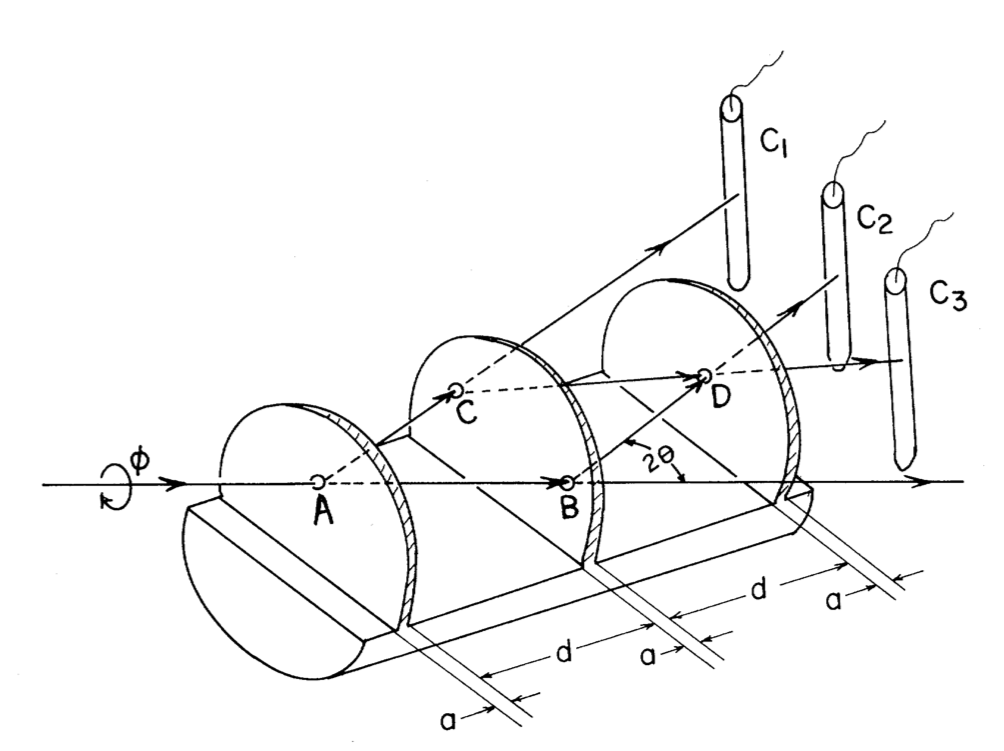}
\includegraphics[scale=0.6]{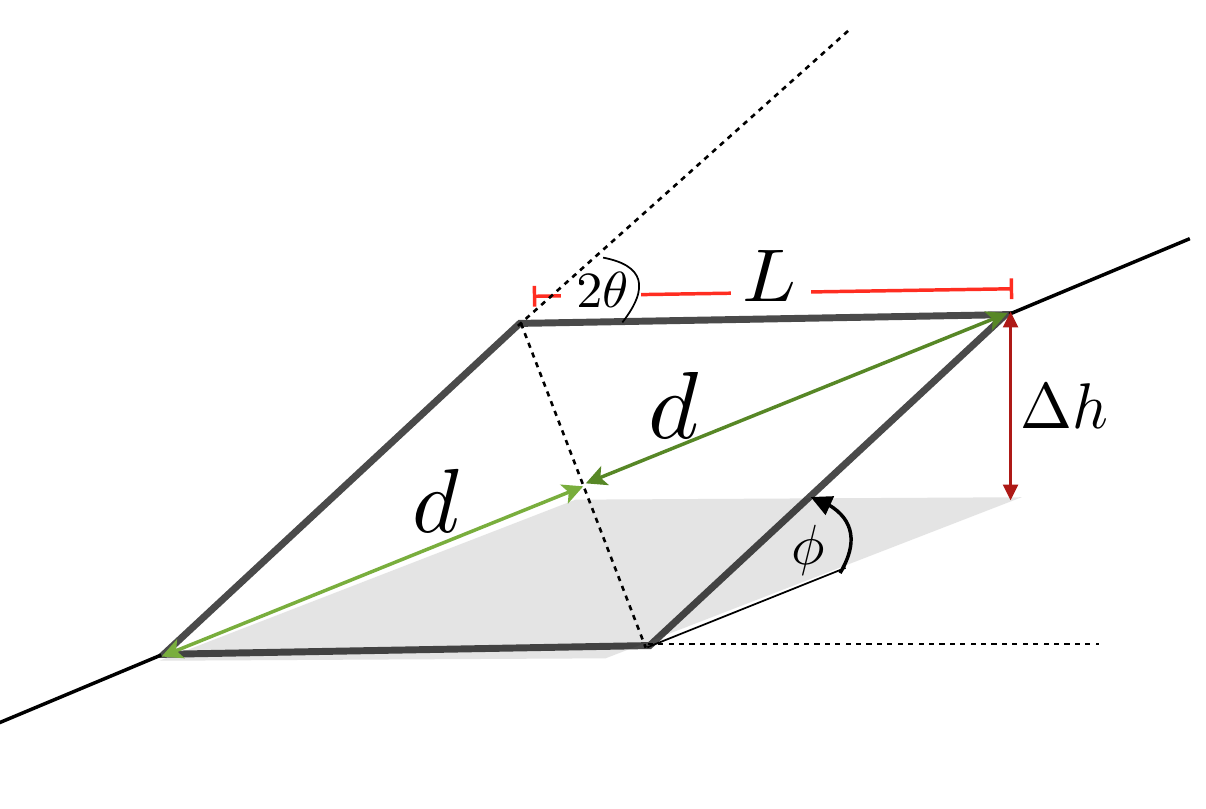}
\caption{\textbf{COW Interferometer}\,| Left: Schematic of the apparatus used in the COW experiment, taken from Ref.~\cite{Colella:1975jc}. The interferometer is rotated about the axis of the first Bragg angle of diffraction. Right: Simplified diagram used to derive the induced phase shift.}
\label{fig:COWsetup}
\end{center}
\end{figure}

The Colella-Overhauser-Werner experiment provided the first evidence of a gravitational effect that is purely quantum mechanical~\cite{Colella:1975jc}. In this experiment, Colella \emph{et al.} used a silicon crystal interferometer to split a beam of neutrons, placing one of the beam paths in a higher gravitational potential (see Figure~\ref{fig:COWsetup}). The difference in the gravitational potential between each arm results in a relative phase shift, which, when recombined, can be measured as modulated intensity. On the length scale of the interferometer, the gravitational field is approximately constant. This allows the relative phase difference between the beams to be calculated using the Wentzel-Kramers-Brillouin (WKB) approximation; that is to integrate the potential difference between the classical trajectories over time~\cite{hall2013quantum}. The two vertical paths of the interferometer contribute phases which cancel out, leaving only the horizontal paths. The phase shift is found to be
\begin{gather}
\Delta\Phi = \frac{2\pi m_\I m_\G g  A \lambda}{h^2}\sin\phi. \tageq\label{COWPHASE}
\end{gather}

A phase shift of this form would be predicted for a quantum mechanical particle in the presence of any scalar potential; in this case, it is the Newtonian gravitational potential. A full description of this effect requires only regular quantum mechanics and Newtonian theory, needing no metric description of gravity, but being unexplainable by classical Newtonian gravity alone. It represents the first evidence of gravity interacting in a truly quantum mechanical way. However, from the perspective of quantum theory, this effect is well understood as a scalar Aharanov-Bohm effect and manifests similarly for charged particles in electric potentials \cite{Allman:1999fh,Zych:2011jz}.

\section{Single slit diffraction in a Newtonian gravitational potential}\label{App:PathIntGravity}

\subsection{Derivation of the propagator}\label{app:PropDeriv}

First consider the Lagrangian for a free particle $L = \frac12m\dot{x}^2$. The Feynman propagator is given by
\begin{align}
\ip{x\pri,t\pri}{x,t} = K_0(x\pri,t\pri;x,t) &= \bigintsss\D \left(x(t)\right) \exp\left[\frac{i}{\hbar}\int_t^{t\pri}\frac12 m \dot{x}(s)^2 ds\right]\tageq\label{eq:FreeFInt}
\n
&= \frac{\exp\left[\frac{im}{2\hbar}\frac{(x\pri - x)^2}{t\pri - t}\right]}{\sqrt{2\pi i \hbar (t\pri - t)/m}}\tageq\label{eq:FreeProp}.
\end{align}
This result will be needed when we consider the propagator for a particle in a linear gravitational potential. In this case the Lagrangian is given by $L = \frac12 m \dot{x}^2 - m g x$, which gives the Feynman propagator
\begin{gather}
K_g(x\pri,t\pri;x,t) = \bigintsss\D \left(x(t)\right) \exp\left[\frac{i}{\hbar}\int_t^{t\pri} ds \left\{ \frac12 m \dot{x}(s)^2  - m g x(s)\right\}\right].\tageq\label{eq:GFP-1}
\end{gather}
To simplify this calculation we express the path $x(t)$ in terms of deviations from the classical trajectory $x_c(t)$ which satisfies the Euler-Lagrange equations of motion. The Feynman measure which sums over all possible paths then becomes a sum over all possible deviations from the classical path. The action expressed in terms of this new parametrisation is
\begin{align}
S[x_c(t) + \delta x(t)] =& \int_{t}^{t\pri} dt \left\{ \frac12 m \Big(\dot{x} +  \delta \dot{x}\Big)^2 - mg\Big(x_c + \delta x\Big) \right\} \n
=&\int_{t}^{t\pri} dt \left\{
\frac12m\dot{x}_c^2 - m g x_c + m\dot{x}_c\delta \dot{x} 
+  \frac12 m(\delta\dot{x})^2 -  mg \delta x \right\},
\end{align}
where $\frac12m\dot{x}_c^2 - m g x_c = \ts s[x_c(t)]$, which is evidently the extremised action given by the classical trajectory. The term containing $\dot{x}_c\delta\dot{x}$ can be integrated by parts, realising that the deviations are zero at the endpoints of the path:
\begin{gather}
S[x_c(t) + \delta x(t)] = S[x_c(t)]  + \int_{t}^{t\pri} \frac12 m (\delta \dot{x})^2 \, dt 
+ [ \ddot{x} \delta x]_{t}^{t\pri}
- \int_{t}^{t\pri}m\delta x \left(\ddot{x}_c + g\right)  dt.
\end{gather}
The last two terms vanish ($\ddot{x}_c = -g$) and substituting the rest into Eq.~\eqref{eq:GFP-1} and factoring out the classical action we arrive at
\begin{gather}
K_g(x\pri,t\pri;x,t) = \exp\left[\frac{i}{\hbar} S\left[x_c(t)\right]\right]\bigintsss\D \left(\delta x(t)\right) \exp\left[\frac{i}{\hbar}\int_t^{t\pri}\frac12 m \delta \dot{x}(s)^2ds\right].\tageq\label{eq:GFP-2}
\end{gather}
The remaining Feynman integral over the deviations is recognised as the free particle propagator in Eq.~\eqref{eq:FreeFInt}, but with the subtle difference being that $x=x\pri=0$. Substituting the integral with the expression from Eq.~\eqref{eq:FreeProp} the propagator becomes
\begin{gather}
K_g(x\pri,t\pri;x,t)= \frac{\exp\left[\frac{i}{\hbar} S\left[x_c(t)\right]\right]}{\sqrt{2\pi i \hbar (t\pri - t)/m}}.\tageq\label{eq:GFP-3}
\end{gather}
Now, all that remains is to find the explicit form for the classical action. We begin with the classical equation of motion, $\ddot{x}_c = -g$, and solve to find the general solution,
\begin{gather}
x_c(t)= -\frac12 g t^2 + a t + b.
\end{gather}
We now impose the boundary conditions $x(t)= x$ and $x(t\pri)= x\pri$ and solve for the constants $a$ and $b$:
\begin{gather}
x = -\frac12 g t^2 + at + b \quad\mbox{and}\quad
x\pri = -\frac12gt\prisq + a t\pri + b\tageq\label{eq:con2}.
\end{gather}
Solving for $a$ and $b$ gives
\begin{gather}
a= \frac{x\pri-x}{t\pri-t} - \frac{g(t^2 - t\prisq)}{2(t\pri - t)}
= \frac{x\pri-x}{t\pri-t} + \frac12 g(t + t\pri),\tageq\label{a}
\end{gather}
and
\begin{align}
b&= \frac12\left(x\pri + x + \frac12 g (t^2 + t\prisq) - a(t+t\pri)\right)\n
&= \frac12\left(x\pri + x - (x\pri - x)\frac{t + t\pri}{t\pri -t} + \frac12 g (t^2 + t\prisq) -\frac12 g (t+t\pri)^2\right)\n
&= \frac12\left(\frac{(x\pri + x)(t\pri-t) - (x\pri - x)(t + t\pri)}{t\pri -t} - gtt\pri\right)\n
&= \frac12\left(\frac{2(xt\pri - x\pri t)}{t\pri -t} -  gtt\pri\right)
= \frac{x t\pri - x\pri t}{t\pri -t} - \frac12 g t t\pri.\tageq\label{b}
\end{align}
Thus, the action of the path taken from $(x,t)$ to $(x\pri, t\pri)$ is
\begin{align}
S[x_c(t)] &= \int_{t}^{t\pri} dt \left\{ \frac12m(-gt + a)^2 - m g(-\frac12 g t^2 + at + b) \right\} \n
&= \frac{m}{2}\int_{t}^{t\pri} dt \left\{2g^2t^2 -2 g a t+ a^2 - 2g b \right\}\n
&= \frac{m}{2}\left(a^2(t\pri-t) - 2g\left(\rule{0cm}{.4cm}a(t\pri-t) + b\right)(t\pri-t) + \frac{2 g^2}{3}(t\pri-t\pri)^3 \right)\\
&= \frac{m}{2}\left\{\frac{(x\pri - x)^2}{t\pri-t} - g(x+x\pri)(t\pri-t) - \frac{g^2}{12}(t\pri-t)^3\right\}.\tageq\label{eq:clAc}
\end{align}
Finally, substituting this into Eq.~\eqref{eq:GFP-3}, we arrive at the complete expression for the propagator for a particle in a gravitational potential,
\begin{align}
K_g(x\pri,t\pri;x,t)  &=\; \frac{\exp\left[\frac{im}{2\hbar}\left\{\frac{(x\pri - x)^2}{t\pri-t} - g(x+x\pri)(t\pri-t) - \frac{g^2}{12}(t\pri-t)^3\right\}\right]}{\sqrt{2\pi i \hbar (t\pri - t)/m}}\tageq\label{eq:GFP}\;.
\end{align}

\subsection{Calculating the single slit wavefunction}\label{sec:SlitWF}

We now consider applying this propagator to the problem at hand. Let's begin by assuming that the slits are long enough to ignore diffraction effects in the $y$ direction. Consider a source of particles at the origin $(0,0)$ and let a double slit be located at distance $z=D$ metres from the source. Each slit has width $2a$ with centre located at $x=\pm b$. The screen is then a further $L$ metres away from the slits. The two dimensional propagator required for this problem is given by a free particle propagator in the $z$-direction multiplied by the gravitational propagator for the $x$ direction as calculated in Eq.~\eqref{eq:GgProp}. This propagator allows us to ask the question of \emph{If a particle initially starts at position $\vr = (0,0)$, what is the probability of finding the it at position $\vrp=(x,D+L)$ on the screen?} This distribution in $x$ will be the the two slit interference pattern that we seek. When computing this amplitude we consider a semi-classical approach. We assume that the `trajectory' of the neutron can be separated into two parts: (a) the path from the source to the slits, followed by (b) the path from the slits to the screen. Quantum mechanically the particles need not pass through the slits and there even exists the possibility of them passing through the slits multiple times before hitting the screen. That being said the probabilities associated with these events are negligible. 

The semi-classical approach is valid provided that the majority of the particle's momentum is in the $z$ direction, such that the wavelength is approximately the $z$-direction wavelength, $\lambda \approx \frac{2\pi \hbar}{m v_z}$. We assume that this wavelength is much smaller than the $z$-direction scale lengths $D$ and $L$ in conjunction with the assumption that these are much larger than the $x$ direction scale lengths. This allows us to consider the particles motion in the $z$-direction as approximately classical and allows for the motion to be partitioned about the slits. The specific propagator $K_g^{(1)}(x,T+\tau;0,0)$, for the process of starting at point $(x=0,z=0)$ at time $t=0$, passing through position $(\w\in[b-a,b+a],D)$ at time $T$ and then arriving at position $(x,D+L)$ at time $T+\tau$ will simply be a product of propagator for each independent component of the path, integrated over the slit distribution $\Omega(\w)$,
\begin{align}
\Omega(\w) &= \left\{\begin{array}{cc}
 1 & b-a < \w<b+a \\
 0 & \text{otherwise}
\end{array}\right.,
\end{align}
\begin{align}
K_g^{(1)}(x,T+\tau;0,0) =& K_0(D,T;0,0)K_0(D+L,T+\tau;D,T) \notag\n
& \quad \quad \times \int_{b-a}^{b+a} K_g(\w,T;0,0)K_g(x,T+\tau;\w,T)d\w\tageq\label{eq:G1Slit-1}.
\end{align}
Evidently for any particular choice of $D$ and $L$, the two $z$ propagators will only give global phase which is identical for all $x$. This global phase has no measurable effects, allowing us to discard the $z$ propagators. This integral is performed in below giving the result
\begin{align}
& K_g^{(1)}(x,T+\tau;0,0) =  \frac{e^{i\phi(x)}} {i\sqrt{2 \lambda (D+L)}} \notag \n
& \hspace{4cm} \times 
\bigg\{C[\sigma_+(x)] - C[\sigma_-(x)] + iS[\sigma_+(x)]-iS[\sigma_-(x)]\bigg\},\tageq\label{eq:G1Slit}
\end{align}
where $C[u] \equiv \int_0^u \cos\left(\frac{\pi}{2}x^2\right)dx$ is the Fresnel cosine function, $S[u] \equiv \int_0^u \sin\left(\frac{\pi}{2}x^2\right)dx$ is the Fresnel sine function and $\eta = 1 + \frac{L}{D}$ and
\begin{align}
&\sigma_{\pm}(x) = \sqrt{\frac{2}{\lambda L}\eta}\left\{(b\pm a) -\frac{x}{\eta} - \frac12g \frac{m^2 \lambda^2}{h^2}DL\right\},\tageq\label{eq:sigmapm2}\n
&\phi(x) = \pi\left\{\frac{x^2}{\lambda(D+L)} - m g x\frac{\lambda(D+L)}{h^2} - \frac{g^2}{12}\frac{m^4\lambda^3}{h^4}(D+L)(D-L)^2 \right\}.\tageq\label{eq:phix2}
\end{align}

The propagator obtains its name for good reason. An initial wavefunction $\psi_0(x)$ convoluted with the propagator will give the future state of the wavefunction for all time $\psi(x,t) = \int G(x,t;s,0) \psi_0(s)ds$. For the purposes of this calculation we can assume a point source of particles such that the initial spacial distribution of the particle is a $\delta$-function. This however means that the wavefunction is the `square root of a $\delta$-function', which is not guaranteed to be defined. That aside, we can calculate the spatial distribution of the particle at the screen, but in order to have this distribution be normalised, we must account for the fact that a large portion of the wavefunction does not pass through the slit. Thus in actual fact the distribution at the screen is given by the conditional probability to be at position $x$ and time $T+\tau$ given that it was at position $x\pri \in [-a,a]$ at time $T$. Fortunately as outlined below, the normalised wavefunction is simply the propagator in Eq.~\eqref{eq:G1Slit} multiplied by a factor $\sqrt{\frac{\lambda D}{2a}}$. Finally we arrive at the normalised wavefunction at the screen, for a particle passing through a single slit of width $2a$, centred at $x=b$,
\begin{align}
\psi^\text{(1)}(x) &= \frac{e^{i\phi(x)}}{i2\sqrt{\eta a}} \bigg\{C[\sigma_+(x)] - C[\sigma_-(x)] + iS[\sigma_+(x)]-iS[\sigma_-(x)]\bigg\},\tageq\label{eq:psi12}
\end{align}
with $\phi(x)$ and $\sigma_\pm(x)$ defined in Eqs.~\eqref{eq:phix} and \eqref{eq:sigmapm}, and $\eta = 1+L/D$. The square of this wavefunction will give the probability distribution for the particle at the slit, which is plotted in Figure~\ref{fig:FallingDiff} for various distances between slit and screen. The pattern clearly appears to shift towards the negative $x$ direction as the screen is moved further from the slit. 

\subsubsection*{Integrating over the slit profile} \label{app:SlitIntegral}

The propagator to arrive at $x$ having passed through a single slit of width $2a$ with centre at $x=b$ is found by integrating over the slit distribution, $\Omega(\w)$, which is $1$ for $b-a < \w<b+a$ and 0 otherwise:
\begin{align}
K_g^{(1)}(x;a,b) =& \int_{-\infty}^\infty A(x,\w)\Omega(\w)d\w  \n
=& \int_{b-a}^{b+a} K_g(\w,T;0,0)K_g(x,T+\tau;\w,T)d\w\n
=&\int_{-\infty}^\infty d\w \, \Omega(\w) \sqrt{\frac{m}{2\pi i \hbar T}} \sqrt{\frac{m}{2\pi i \hbar \tau}} 
\exp\left[\frac{im}{2\hbar}\left\{\frac{\w^2}{T} - g\w T - \frac{g^2}{12}T^3\right\}\right]\\
\notag
&\qquad \times \exp\left[\frac{im}{2\hbar}\left\{\frac{(x-\w)^2}{\tau} - g(x+\w) \tau - \frac{g^2}{12}\tau^3\right\}\right].
\tageq\label{eq:G1Slit-3}
\end{align}
Completing the square in $\w$,
\begin{align}
 \frac{\w^2}{T} - g\w T - & \frac{g^2}{12}T^3 + \frac{(x-\w)^2}{\tau} - g(x+\w) \tau - \frac{g^2}{12}\tau^3\n
&= \frac{\w^2}{T} + \frac{\w^2}{\tau} - \frac{2 x \w}{\tau} - g\w(T + \tau) +\frac{x^2}{\tau} - gx\tau +\frac{g^2}{4}(T^3 + \tau^3)\n
&= \w^2\frac{T + \tau}{T\tau} - 2\w\left(\frac{x}{\tau} + \frac12g(T+\tau)\right) +\frac{x^2}{\tau}  - gx\tau - \frac{g^2}{12}(T^3 + \tau^3)\n
&=\zeta\left(\w - \frac{\kappa}{\zeta}\right)^2 - \frac{\kappa^2}{\zeta} +\frac{x^2}{\tau}  - gx\tau - \frac{g^2}{12}(T^3 + \tau^3),
\end{align}
where $\zeta = \frac{T + \tau}{T\tau}$ and $\kappa = \frac{x}{\tau} + \frac12g(T+\tau)$. Returning to Eq.~\eqref{eq:G1Slit-3},
\begin{align}
K_g^{(1)}(x;a,b) &=  e^{i\phi(x,T,\tau)}\sqrt{\frac{m}{2\pi i \hbar T}} \sqrt{\frac{m}{2\pi i \hbar \tau}} \int_{-\infty}^\infty d\w \exp\left[\frac{im\zeta}{2\hbar}\left(\w - \frac{\kappa}{\zeta}\right)^2\right],
\end{align}
where $\phi(x,T,\tau) = \frac{m}{2\hbar}\left(\frac{x^2}{\tau} -\frac{\kappa^2}{\zeta} - gx\tau - \frac{g^2}{12}(T^3 + \tau^3)\right)$ is the phase produced by terms not dependent on $\w$. We make the substitution $\ts v = \sqrt{\frac{m\zeta}{\pi \hbar}}(\w - \frac{\kappa}{\zeta})$, and define new limits of integration $\sigma_{\pm}(x) =\sqrt{\frac{m\zeta}{\pi \hbar}}\left((b\pm a) - x \frac{T}{T+\tau} - \frac12gT\tau\right)$:
\begin{align}
K_g^{(1)}(x;a,b) &=e^{i\phi(x,T,\tau)}\sqrt{\frac{2 \pi i  \hbar}{m\zeta}}\sqrt{\frac{m^2}{(2\pi i \hbar)^2 T\tau}}\int_{\sigma_-}^{\sigma_+}\exp\left[\frac{i\pi}{2} v^2\right]dv \n
&= \frac{e^{i\phi(x,T,\tau)}}{\sqrt{(2i)^2 \pi \hbar (T+\tau)/m}} \int_{\sigma_-}^{\sigma_+}\left\{\cos\left(\frac{i\pi}{2}v^2 \right)+ i \sin\left(\frac{i\pi}{2}v^2\right)\right\}\, dv\n
&= \frac{e^{i\phi(x,T,\tau)}}{2i\sqrt{\pi \hbar (T+\tau)/m}}\bigg\{C[\sigma_+(x)] - C[\sigma_-(x)] + iS[\sigma_+(x)]-iS[\sigma_-(x)]\bigg\},\tageq\label{eq:G1Slit-4}
\end{align}
where $C[u] \equiv \int_0^u \cos\left(\frac{\pi}{2}x^2\right)dx$ is the Fresnel cosine function, $S[u] \equiv \int_0^u \sin\left(\frac{\pi}{2}x^2\right)dx$ is the Fresnel sine function. Now to simplify $\phi(x,T,\tau)$ we first have
\begin{align}
\kappa^2 &= \frac{x^2}{\tau^2} + \frac{x}{\tau}g(T+\tau) - \frac{g^2}{12}(T+\tau)^2,\n
\frac{\kappa^2}{\zeta} &= \frac{x^2}{\tau^2}\frac{T \tau}{T+\tau} + \frac{x}{\tau}g(T+\tau)\frac{T \tau}{T+\tau} - \frac{g^2}{12}(T+\tau)^2\frac{T \tau}{T+\tau}\n
&= \frac{x^2 T}{\tau(T+\tau)} + gxT - \frac{g^2}{12}T \tau(T+\tau)
\end{align}
to get
\begin{align}
\phi(x,T,\tau)&= \frac{m}{2\hbar}\left\{\frac{x^2}{\tau} -\frac{\kappa^2}{\zeta} - gx\tau - \frac{g^2}{12}(T^3 + \tau^3)\right\}\n
&=\frac{m}{2}\left\{\frac{x^2(T+\tau) - x^2 T}{\tau (T+\tau)} - gx(T+\tau) - \frac{g^2}{12}(T^3 + \tau^3 - T\tau(T+\tau))\right\}\n
&= \frac{m}{2}\left\{\frac{x^2}{T +\tau} - gx(T+\tau)  - \frac{g^2}{12}(T+\tau)(T-\tau)^2\right\}.
\end{align}
We can make use of the approximation $v_z\gg v_x$ and that $\lambda \approx \frac{h}{mv_z}$ to find expressions $T=\frac{m \lambda D}{h}$ and $\tau=\frac{m\lambda L}{h}$. Thus, we have 
\begin{gather}
T \pm \tau = \frac{m\lambda(D\pm L)}{h}
\quad \mbox{and} \quad
T\tau = \frac{m^2\lambda^2}{h^2}DL.
\end{gather}
Using these we get
\begin{gather}
\frac{m\zeta}{\pi \hbar}= \frac{2}{\lambda}\left(\frac{1}{D} + \frac{1}{L}\right) \qquad \mbox{where} \qquad
\zeta = \frac{T+\tau}{T\tau} = \frac{h}{m\lambda}\frac{D+L}{DL} =\frac{h}{m\lambda}\left(\frac{1}{D} + \frac{1}{L}\right).
\end{gather}
Next, let
\begin{gather}
\eta = \frac{T}{T+\tau} = \frac{D}{D+L} = \frac{1}{1+L/D},
\end{gather}
allowing us to express $\phi$ and $\sigma_{\pm}$ in terms of $L$, $D$ and $\lambda$:
\begin{align}
\notag
&\sigma_{\pm}(x) =\sqrt{\frac{m\zeta}{\pi \hbar}}\left((b\pm a) - x \frac{T}{T+\tau} - \frac12gT\tau\right) \notag\n
&\phantom{\sigma_{\pm}(x)}
= \sqrt{\frac{2}{\lambda L}\eta}\left\{(b\pm a) -\frac{x}{\eta} - \frac12g \frac{m^2 \lambda^2}{h^2}DL\right\}, \n
&\phi(x,T,\tau) = \frac{m}{2}\left\{\frac{x^2}{T +\tau} - gx(T+\tau)  - \frac{g^2}{12}(T+\tau)(T-\tau)^2\right\},\n
&\phi(x)=\pi\left\{\frac{x^2}{\lambda(D+L)} - m g x\frac{\lambda(D+L)}{h^2} - \frac{g^2}{12}\frac{m^4\lambda^3}{h^4}(D+L)(D-L)^2 \right\}. 
\end{align}
This is the form of the propagator given in Eq.~\eqref{eq:G1Slit}.

\subsubsection*{Normalisation of the distribution at the screen}\label{app:Re-normalisation}

As derived in the first Section the propagator for the process of starting at position $\vr = (0,0)$, passing through the point $(x\pri\in[b-a,b+a],D)$ and finally being detected at position $\vrp=(x,D+L)$ on the screen is not the same as the wave-function at the screen. To obtain this we must first convolve the propagator with a initial wavefunction whose square magnitude is a $\delta$-function giving the wavefunction as seen at the other side of the slit. This wavefunction however will not be normalised due to the fact that only a portion of the initially normalise wavefunction has been propagated beyond the slits. It can be renormalised however by scaling by the probability of passing through the slit. Unfortunately the `square root of a $\delta$-function' is not always well defined as is the case for operators acting on any distribution. We can however attempt to use a Gaussian with variance $\sigma$ as the initial wavefunction, compute the quantity of interest and take the limit $\sigma \rightarrow 0$. Under suitable circumstances the limit will be defined giving the desired result.

We will begin with an initial wavefunction that is the square root of a Gaussian
\begin{align}
\psi_\sigma(x) = g_\sigma(x) &= \frac{1}{(2\pi \sigma^2)^{\frac14}} e^{-\frac{x^2}{4\sigma^2}}\tageq\label{Gaussianwavefunction}.
\end{align}
However we notice that the square root of a Gaussian is simply another Gaussian of variance $\rho =\sigma\sqrt{2}$ multiplied by the factor $(8\pi \sigma^2)^\frac{1}{4}$. So the initial function can represented as
\begin{align}
\psi_\rho(x) &= (4\pi \rho^2)^\frac{1}{4} \frac{e^{-\frac{x^2}{2\rho^2}}}{\sqrt{2\pi \rho^2}} = (4\pi \rho^2)^\frac{1}{4} g_\rho(x)\tageq\label{redefinedGaussianWF}.
\end{align}

Convolving this with the propagator $K_g^{(1)}(x,T+\tau;x_0,0)$ will give the un-normalised wavefunction at the screen:
\begin{gather}
\psi_\rho(x,T+\tau) = \int_{\infty}^\infty dx_0 K_g^{(1)}(x,T+\tau;x_0,0)\psi_\rho(x_0).\tageq \label{IntegrateInitialWF1}
\end{gather}
To renormalise this, we scale by the probability of the particle passing through the slit. The probability of the particle being in $x\in[b-a,b+a]$ at time $T$ is 
\begin{gather}
P(x\in[b-a,b+a];T) = \int_{b-a}^{b+a}  \, \left|\int_{-\infty}^{\infty}K_g(x\pri,T;x_0,0)\psi_\rho(x_0)dx_0\right|^2dx\pri,
\end{gather}
which gives that the renormalised wavefunction at the screen is
\begin{align}
\psi_\rho\pri(x,T+\tau) &= \frac{\psi(x,T+\tau)}{\sqrt{\int_{-a}^a \, |\int_{-\infty}^{\infty}K_g(x\pri,T;x_0,0)\psi_\rho(x_0)dx_0|^2dx\pri }}\n
&=\frac{\int_{\infty}^\infty  K_g^{(1)}(x,T+\tau;x_0,0)\psi_\rho(x_0)dx_0}{\sqrt{\int_{b-a}^{b+a} \, |\int_{-\infty}^{\infty}K_g(x\pri,T;x_0,0)\psi_\rho(x_0)dx_0|^2dx\pri }}\n
&= \frac{(4\pi \rho^2)^\frac{1}{4}\int_{\infty}^\infty  K_g^{(1)}(x,T+\tau;x_0,0)g_\rho(x_0)dx_0}{(4\pi \rho^2)^\frac{1}{4}\sqrt{\int_{b-a}^{b+a}  \, |\int_{-\infty}^{\infty}K_g(x\pri,T;x_0,0)g_\rho(x_0)dx_0|^2dx\pri}}.
\end{align}
Now the limit $\sigma\rightarrow 0$ can equivalently be taken as $\rho\rightarrow 0$. The Gaussians $g_\rho(x)$ then become delta functions $\delta(x)$:
\begin{align}
\psi\pri(x,T+\tau)&= \lim_{\rho-\rightarrow 0}\frac{\int_{\infty}^\infty  K_g^{(1)}(x,T+\tau;x_0,0)g_\rho(x_0)dx_0}{\sqrt{\int_{b-a}^{b+a}  \, |\int_{-\infty}^{\infty}K_g(x\pri,T;x_0,0)g_\rho(x_0)dx_0|^2dx\pri}}\n
&= \frac{K_g^{(1)}(x,T+\tau;0,0)}{\sqrt{\int_{b-a}^{b+a}  \, |K_g(x\pri,T;0,0)|^2dx\pri}}
= \frac{K_g^{(1)}(x,T+\tau;0,0)}{\sqrt{\int_{b-a}^{b+a}  \,(2\pi \hbar T/m)^{-1}dx\pri}}\n
&= K_g^{(1)}(x,T+\tau;0,0) \sqrt{\frac{\frac{h}{m} T}{2a}}
= K_g^{(1)}(x,T+\tau;0,0) \sqrt{\frac{\frac{m\lambda D}{h} \frac{h}{m}}{2a}}\n
&= K_g^{(1)}(x,T+\tau;0,0) \sqrt{\frac{\lambda D}{2a}},
\end{align}
where $|K_g(x\pri,T;0,0)|^2$ was taken from Eq.~\eqref{eq:GFP}. Thus we see that the normalised wavefunction at the screen is given by multiplying the propagator by the factor $\sqrt{\frac{\lambda D}{2a}}$.

\section{Tracing out spin from a matrix propagator}\label{app:PropagatorTrace}

This is best achieved using the density operator prescription. The pure density operator $\rho$ for a quantum state $\ket{\psi}$ is $\rho = \ket{\psi}\bra{\psi}$. For a state comprised of two subsystems, we can ignore the state of a subsystem by tracing it out. This is given by the operation $\tr_B[X_{AB}] = \sum_k \bra{k}_B X_{AB} \ket{k}_B$, where $X_{AB}$ is an operator on the composite system $AB$ and $\{\ket{k}_B\}$ forms a complete basis for subsystem $B$. 

Here, we would like to trace out the spin state. The initial  density operator is $\rho_0 = \out{\chi_0}\otimes\out{\psi_0}$, where $\ip{x}{\psi_0} = \psi_0(x)$, is the spatial distribution of the particle, and it is assumed that, initially, the spatial location of the particle is uncorrelated with the spin state. The state of the system at later time $t$ is given by $\rho(t) = U(t)\rho_0U^\dagger(t)$.
We can represent the time evolution operator in terms of the propagator by making use of the resolution of the identity $\sum_{\{\chi,\chi\pri\}\in\{\darr,\uarr\}}\int dx \,dx\pri \kb{x\pri,\chi\pri}{x,\chi} = \id$:
\begin{gather}
U(t)=\sum_{\{\chi,\chi\pri\}\in\{\darr,\uarr\}}\int dx \,dx \pri K_g^{\chi\pri\!,\chi}(x\pri,t;x,0) \kb{x\pri,\chi\pri}{x,\chi}\tageq\label{eq:UnitaryProp},
\end{gather}
which gives that,
\begin{align}
\rho(t) =& \sum \int dx \, dx\pri dy\, dy\pri K_g^{\chi\pri\!,\chi}(x\pri,t;x,0)K_g^{^*\phi\pri\!,\phi}(y\pri,t;y,0)\ket{x\pri,\chi\pri}\notag\n 
& \qquad \times
\bk{\vphantom{\chi\pri}x,\chi}{\rho_0}{\vphantom{\chi\pri}y,\phi}\bra{y\pri,\phi\pri},
\end{align}
where the sum is over all spin variables. We can then take the trace over the spin subspace to give,
\begin{align}
\tilde{\rho}(t) = \tr_\text{spin}[\rho(t)] =& \sum \int dx \, dx\pri dy\, dy\pri K_g^{\chi\pri\!,\chi}(x\pri,t;x,0)K_g^{^*\phi\pri\!,\phi}(y\pri,t;y,0)
\notag\n
& \qquad \times\bk{\vphantom{\chi\pri}x,\chi}{\rho_0}{\vphantom{\chi\pri}y,\phi} \ip{\chi\pri}{\phi\pri} \kb{x\pri}{y\pri}.
\end{align}
The spatial probability distribution is then given by the expectation of the position operator $\exx{\hat{x}} = \tr[\tilde{\rho}(t)\hat{x}]$.  Making use of the fact that $\ip{\chi\pri}{\phi\pri} = \delta_{\chi\pri,\phi\pri}$ and $\ip{y\pri}{x\pri} = \delta(x\pri - y\pri)$, we find
\begin{gather}
\exx{\hat{x}} = \sum \int dx \, dy K_g^{\chi\pri\!,\chi}(x\pri,t;x,0)K_g^{^*\chi\pri\!,\phi}(x\pri,t;y,0)\bk{\vphantom{\chi\pri}x,\chi}{\rho_0}{\vphantom{\chi\pri}y,\phi}.\tageq\label{eq:exofx}
\end{gather}
At this point, we work with the term $\bk{\vphantom{\chi\pri}x,\chi}{\rho_0}{\vphantom{\chi\pri}y,\phi}$
\begin{align}
\bk{\vphantom{\chi\pri}x,\chi}{\rho_0}{\vphantom{\chi\pri}y,\phi} =& \bk{\vphantom{\chi\pri}x,\chi}{\Big(\out{\chi_0}\otimes\out{\psi_0}\Big)}{\vphantom{\chi\pri}y,\phi}\n
=& \bra{\chi}{\Big(\alpha\ket{\uarr} + \beta\ket{\darr}\Big)\Big(\alpha^*\bra{\uarr} + \beta^*\bra{\darr}\Big)}\ket{\phi}\ip{x}{\psi_0}\ip{\psi_0}{y}\n
=& \bra{\chi}{\Big(|\alpha|^2\out{\uarr} + |\beta|^2\out{\darr} + \alpha^*\beta\kb{\darr}{\uarr} + \beta*\alpha\kb{\uarr}{\darr}\Big)}\ket{\phi}
\tageq\label{eq:ProjectionInitialRho}\n\notag
& \quad \times \psi_0(x)\psi_0^*(y).
\end{align}
Since the matrix propagator in Eq.~\eqref{eq:DefMatrixProp2} is diagonal, we can immediately discard the $\kb{\darr}{\uarr}$ and $\kb{\uarr}{\darr}$ terms when substituting into the expression for the spatial distribution in Eq.~\eqref{eq:exofx},
\begin{align}
\exx{\hat{x}} =& \int dx \, dy |\alpha|^2K_g^{\uarr,\uarr}(x\pri,t;x,0)\psi_0(x)\left(K_g^{\uarr,\uarr}(x\pri,t;y,0)\psi_0(y)\right)^*\\
 & + |\beta|^2K_g^{\darr,\darr}(x\pri,t;x,0)\psi_0(x)\left(K_g^{\darr,\darr}(x\pri,t;y,0)\psi_0(y)\right)^*
\notag\n
=& |\alpha|^2\left|\int dx K_g^{\uarr,\uarr}(x\pri,t;x,0)\psi_0(x)\right|^2 +|\beta|^2\left|\int dx K_g^{\darr,\darr}(x\pri,t;x,0)\psi_0(x)\right|^2. \tageq\label{appeq:ConvexSum}
\end{align}
This result is simply a convex sum of the initial spatial distribution evolved by each propagator. 

\bibliography{ResearchRef}
\bibliographystyle{ieeetr}

\end{document}